\documentclass[english,aps,groupedaddress,superscriptaddress]{revtex4}
\usepackage[T1]{fontenc}
\usepackage[latin9]{inputenc}
\usepackage{float}
\usepackage{amsthm}
\usepackage{amsmath}
\usepackage{graphicx}
\usepackage{esint}

\makeatletter

\providecommand{\tabularnewline}{\\}

\@ifundefined{textcolor}{}
{%
 \definecolor{BLACK}{gray}{0}
 \definecolor{WHITE}{gray}{1}
 \definecolor{RED}{rgb}{1,0,0}
 \definecolor{GREEN}{rgb}{0,1,0}
 \definecolor{BLUE}{rgb}{0,0,1}
 \definecolor{CYAN}{cmyk}{1,0,0,0}
 \definecolor{MAGENTA}{cmyk}{0,1,0,0}
 \definecolor{YELLOW}{cmyk}{0,0,1,0}
}

\usepackage[]{units}
\providecommand{\e}[1]{\ensuremath{\times 10^{#1}}}

\makeatother

\usepackage{babel}
\begin{document}

\title{The limits of statistical significance of Hawkes processes fitted
to financial data}

\author{Mehdi Lallouache}

\affiliation{Chaire de Finance Quantitative, Laboratoire de Mathématiques Appliquées
aux Systèmes, Ecole Centrale Paris, Châtenay-Malabry, 92290, France}

\author{Damien Challet}

\affiliation{Chaire de Finance Quantitative, Laboratoire de Mathématiques Appliquées
aux Systèmes, Ecole Centrale Paris, Châtenay-Malabry, 92290, France}

\affiliation{Encelade Capital SA, EPFL Innovation Park, Bâtiment C, 1015 Lausanne,
Switzerland}
\begin{abstract}
Many fits of Hawkes processes to financial data look rather good but
most of them are not statistically significant. This raises the question
of what part of market dynamics this model is able to account for
exactly. We document the accuracy of such processes as one varies
the time interval of calibration and compare the performance of various
types of kernels made up of sums of exponentials. Because of their
around-the-clock opening times, FX markets are ideally suited to our
aim as they allow us to avoid the complications of the long daily
overnight closures of equity markets. One can achieve statistical
significance according to three simultaneous tests provided that one
uses kernels with two exponentials for fitting an hour at a time,
and two or three exponentials for full days, while longer periods
could not be fitted within statistical satisfaction because of the
non-stationarity of the endogenous process. Fitted timescales are
relatively short and endogeneity factor is high but sub-critical at
about 0.8.
\end{abstract}
\maketitle

\section{Introduction}

Hawkes processes are a natural extension of Poisson processes in which
self-excitation causes event clustering \citep{Hawkes1971,Hawkes1971a}.
Originally applied to the modeling of earthquake occurrences \citep{Ogata1988,Ogata1999},
they have proven to be useful in many fields (e.g. neuroscience, criminology
and social networks modeling \citep{Chornoboy1988,Pernice2012,Mohler2011,Crane2008,Yang2013}).
This is because of their tractability and the ever-increasing number
of estimation methods \citep{Marsan2008,Reynaud-Bouret2010,Lewis2011,Bacry2012,DaFonseca2013,Bacry2014}.
Since many types of financial market events such as mid-quote changes,
extreme return occurrences or order submissions are clustered in time,
Hawkes processes have become a standard tool in finance too.

In the context of market microstructure, Hawkes processes were first
introduced by \citet{Bowsher2007}, who simultaneously analyzed trades
time and mid-quotes changes with a multivariate framework. Two others
pioneer approaches are the ones by \citet{Bauwens2004} and \citet{Hewlett2006}
who focused on the durations between transactions. Subsequently, \citet{Large2007}
supplemented transaction data with limit orders and cancellations
data in a ten-variate Hawkes process in order to measure the resilience
of an London Stock Exchange order book. \citet{Bacry2012a} have recently
modeled the mid-price change as the difference between two Hawkes
processes and showed that the resulting price exhibits microstructure
noise and the Epps effect. \citet{Jaisson2013} established that under
a suitable rescaling a nearly unstable Hawkes process converges to
a Heston model. \citet{Bacry2014a} used an enhanced version of the
model to account for market impact. Finally, \citet{Jedidi2013} modeled
the full order book with a multivariate Hawkes setup and proved that
the resulting price diffuses at large time scales. Remarkably, Hawkes
processes are also applied to other financial topics such as VaR estimation
\citep{Chavez-Demoulin*2005,Chavez-Demoulin2012}, trade-through modeling
\citep{Toke2011}, portfolio credit risk \citep{Errais2010}, or financial
contagion across regions \citep{Ait-Sahalia2010} and across assets
\citep{Bormetti2013}.

It is widely accepted among researchers that only a small fraction
of price movements is directly explained by external news releases
(e.g. \citet{cutler1989moves,joulin2008stock}). Thus, the price dynamics
is mostly driven by internal feedback mechanisms, which corresponds
to what Soros calls \textquotedblleft market reflexivity'' \citep{Soros1987}.
In the framework of Hawkes processes, endogeneity comes from self-excitation
while the baseline activity rate is deemed exogenous (see Sec. \ref{sec:Hawkes-process-and}
for a mathematical definition). In other words, these processes provide
a straightforward way to measure the importance of endogeneity, for
example in the E-mini S\&P futures \citep{Filimonov2012,Hardiman2013a}.
\citet{Filimonov2012} argued that the level of endogeneity has increased
steadily in the last decade due to the advent of high-frequency and
algorithmic trading. \citet{Hardiman2013a}\textit{ }showed that it
is only\textit{ }the\textit{ short-term }endogeneity (linked to increases
of computer power and speed, and, indeed, HFT) that has increased
over the years, while the endogeneity factor has been very stable
and close to 1, the special value at which the process becomes totally
self-referential and unstable. Fitting Hawkes processes to financial
data requires some care: one should not use a single exponential self-excitation
kernel \citep{Hardiman2013a}, while many other biases may affect
fits with long-tailed kernels on long time periods \citep{Filimonov2013}.

Nobody claims that Hawkes process are the exact description of the
whole dynamics of financial markets. However, testing the significance
of the fits is not a current priority in the literature. Given the
fact that the fits are usually visually satisfactory, it seems obvious
that statistical significance may be obtained in some cases. Here,
we wish to assess the extent (and the limits) of the explanatory power
of Hawkes processes with several possibly types of parametric kernels,
according to three statistical tests. One of the difficulties in obtaining
significant fits come from jumps in trading activity such as those
occurring when markets open and close. This is why we work on data
from FX markets which have the advantage of operating continuously
for longer periods. There may still be discontinuities, either implicit
(e.g. fixing time) or explicit (e.g. week-end closures) in our FX
data, but at least one day of FX data spans many more hours than one
day of equity market data and is thus more suitable to our aim. Hence,
\emph{a minori}, one may extrapolate most of our failures to fit correct
Hawkes processes to other types of data with more significant activity
discontinuities.

The two other papers on FX data and Hawkes processes have a different
focus than ours: \citet{Hewlett2006} deals with the relatively illiquid
EUR/PLN currency pair and uses a single-exponential kernel. \citet{Rambaldi2014}
also use EBS data (with the same time resolution as ours) and studies
the dynamics of best quotes around important news. Because our data
set consists of order book snapshots every \unit[0.1]{s} (see Sec.
\ref{sec:data} for more details), we can trace most trades but not
mid price changes. This is why we fit a univariate Hawkes process
to EUR/USD trade arrivals. The endogeneity parameter is then the average
number of trades triggered by a single trade.

The structure of the paper is as follows: we first define Hawkes processes,
the fitting method, the parametric kernels and the statistical tests
that we will use. We first show that Hawkes processes excel at fitting
one hour of FX data, are fairly good for a single day, and fail when
used for two consecutive days.

\section{Hawkes processes \label{sec:Hawkes-process-and}}

An univariate Hawkes process is a linear self-exciting point process
with an intensity given by

\begin{align}
\lambda_{t} & =\mu_{t}+\int_{0}^{t}\phi(t-s)dN_{s}\nonumber \\
 & =\mu_{t}+\sum_{t_{i}<t}\phi(t-t_{i}),\label{eq:HP_def}
\end{align}

where $\mu_{t}$ is a baseline intensity describing the arrival of
exogenous events and the second term is a weighted sum over past events.
The kernel $\phi(t-t_{i})$ describes the impact on the current intensity
of a previous event that took place at time $t_{i}$.

A Hawkes process can be mapped to (and interpreted as) a branching
process, where exogenous ``mother'' events occurring with intensity
$\mu_{t}$ can trigger one or more ``child'' events. In turn, each
of these children, can trigger multiple child events (or ``grand-child''
respectively to the original event), and so on. The quantity $n\equiv\int_{0}^{\infty}\phi(s)ds$
controls the size of the endogenously generated families. Indeed,
$n$ is the \textit{branching ratio} of the process, which is defined
as the average number of children for any event. Therefore, $n$ quantifies
market reflexivity in an elegant way. Three regimes exist depending
on the branching ratio value:
\begin{itemize}
\item a sub-critical regime $(n<1)$ where families dies out almost surely,
\item the critical regime $(n=1)$, where one family lives indefinitely
without exploding. In the language of Hawkes process, this requires
$\mu=0$ to be properly defined and it is equivalent to Hawkes process
without ancestors studied by \citet{Bremaud2001},
\item the explosive regime $(n>1)$, where a single event triggers an infinite
family with a strictly positive probability.
\end{itemize}
Evaluating $n$ gives a simple measure of the market ``distance''
to criticality. For $n\leq1$, the process is stationary if $\mu_{t}$
is constant. In this case, the branching ratio is also equal to the
average proportion of endogenously generated events among all events.

\subsection{Parametric kernels}

We compare the performance of the following kernels, each labeled
by its own index.
\begin{itemize}
\item Sum of exponentials:
\[
\phi_{M}(t)=\sum_{i=1}^{M}\alpha_{i}e^{-t/\tau_{i}},
\]
where $M$ is the number of exponentials. The amplitudes $\alpha_{i}$
and timescales $\tau_{i}$ of the exponentials are the estimated parameters.
The branching ratio is then given by: $n=\sum_{i=1}^{M}\alpha_{i}\tau_{i}=\sum_{i=1}^{M}n_{i}.$
\item Approximations of power-laws have the advantage of needing a few parameters
only. As a consequence, fitting them to data is much easier. Approximate
power-law kernel is given by
\[
\phi_{M}^{{\scriptscriptstyle \text{PL}}}(t)=\frac{n}{Z}\sum_{i=0}^{M-1}a_{i}^{-(1+\epsilon)}e^{-\frac{t}{a_{i}}},
\]
where
\[
a_{i}=\tau_{0}m^{i}.
\]
$M$ controls the range of the approximation and $m$ its precision.
$Z$ is defined such that $\intop_{0}^{\infty}\phi_{PL}(t)dt=n$.
The parameters are the branching ratio $n$, the tail exponent $\epsilon$
and the smallest timescale $\tau_{0}$.
\item Approximate power-law with a short lags cut-off \citep{Hardiman2013a}:
\[
\phi_{M}^{{\scriptscriptstyle \text{HBB}}}(t)=\frac{n}{Z}\left(\sum_{i=0}^{M-1}a_{i}^{-(1+\epsilon)}e^{-\frac{t}{a_{i}}}-Se^{-\frac{t}{a_{-1}}}\right),
\]
the definition is the same as $\phi_{M}^{{\scriptscriptstyle \text{PL}}}$
with the addition of a smooth exponential drop for lags shorter than
$\tau_{0}$. $S$ is defined such that $\phi_{M}^{{\scriptscriptstyle \text{HBB}}}(0)=0$.
\item We propose a new type of kernels, made up of an approximate power-law
$\phi_{M}^{{\scriptscriptstyle \text{PL}}}$ and one exponential with
free parameters. This is to allow for a greater freedom in the structure
of time scales. The kernel is then defined as 
\[
\phi_{M}^{{\scriptscriptstyle \text{PL}_{\text{x}}}}(t)=\frac{n}{Z}\left(\sum_{i=0}^{M-1}a_{i}^{-(1+\epsilon)}e^{-\frac{t}{a_{i}}}+be^{-\frac{t}{\tau}}\right),
\]
where the exponential term adds two parameters $b$ and $\tau$. The
other variables have the same meaning as above.
\end{itemize}
When a kernel is a sum of exponentials, one can exploit a recursive
relation for the log-likelihood calculation that reduces the computational
complexity from $\mathcal{O}(N^{2})$ to $\mathcal{O}(N)$ (see \citet{Ozaki1979}).
It provides reasonable computation time on a single workstation since
N is $\mathcal{O}(10^{4})$. The first form is the most flexible and
can approximate virtually any continuous function, at the cost of
extra-parameters and more sloppiness \citep{Waterfall2006}. The second
and third ones aim to reproduce the long memory observed in many market
but are less flexible; their effective support may span well beyond
the fitting period. The last one tries to combine the best of both
worlds.

Once a kernel form is specified, we use the L-BFGS-B algorithm \citep{Byrd1995}
to estimate the parameters that maximize the log-likehood. For each
fit we try different starting points to avoid local maxima.

Using multivariate Hawkes process to fit the arrival and the reciprocal
influence of buy and sell trades systematically yields null cross-terms.
Both buy and sell trades yield indistinguishable results; we therefore
focus on buy trades.

\subsection{Goodness-of-fits tests\label{sub:Goodness-of-fits-tests}}

The quality of the fits is assessed on the time-deformed series of
durations $\left\{ \theta_{i}\right\} $, defined by

\[
\theta_{i}=\int_{t_{i-1}}^{t_{i}}\hat{\lambda}_{t}dt,
\]
where $\hat{\lambda}$ is the estimated intensity and $\left\{ t_{i}\right\} $
are the empirical timestamps. If a Hawkes process describes the data
correctly, the $\theta_{i}$s must be (i) independent and (ii) exponentially
distributed with unit rate. The maximum-likelihood estimation, by
construction, tends to maximize the exponential nature of the $\theta$s,
but not their independence. This explains why QQ-plots of the resulting
$\theta$s are visually very satisfying as long as the kernel contains
than more one exponential. 

Visual checks of QQ-plots is only one of the available criteria, many
of them being more precise and rigorous. Indeed, property (i) can
be tested by the Ljung-Box test, which examines the null hypothesis
of absence of auto-correlation in a given time-series. We use here
a slight modification of the original test statistic from \citet{LJUNG1978},
defined as
\[
Q=N(N+2)\sum_{k=2}^{h+1}\frac{\hat{\rho}_{k}^{2}}{n-k},
\]

where $N$ is the sample size, $\hat{\rho}_{k}$ is the sample autocorrelation
at lag $k,$ and $h$ is the number of lags being tested. Under the
null, $Q$ follows a $\chi^{2}$ with $h$ degrees of freedom. Note
that we start the sum at $k=2$ (instead of 1). This is because of
the systematic small one-step anti-correlation introduced by the data
cleaning procedure (Sec. \ref{sub:Treatment}). In other words, we
wish to test the absence of auto-correlation at lags that are unaffected
by this procedure.

Property (ii) is assessed by two tests
\begin{enumerate}
\item Kolmogorov-Smirnov test (KS henceforth), based on the maximal discrepancy
between the empirical cumulative distribution and the exponential
cumulative distribution. The asymptotic distribution under the null
is the Kolmogorov distribution. It is known to be a very (even excessively)
demanding test.
\item \citet{Engle1998} Excess Dispersion test (ED henceforth), which verifies
the lack of excess dispersion in the residuals. The test statistic
reads: 
\[
S=\sqrt{N}\frac{\hat{\sigma}^{2}-1}{\sqrt{8}},
\]
 where $\hat{\sigma}^{2}$ is the sample variance of $\theta$ which
should be equal to $1.$ Under the null, $S$ has a limiting normal
distribution.
\end{enumerate}
All these three tests check basic but essential properties of the
$\theta$s.

\section{data\label{sec:data}}

\subsection{Description}

We study EUR/USD inter-dealer trading from January 1, 2012 to March
31, 2012. The data comes from EBS, the leading electronic trading
platform for this currency pair. A message is recorded every \unit[0.1]{s}.
It contains the highest buying deal price and the lowest selling deal
price with the dealt volumes, as well as the total signed volume of
trades in the time-slice. Orders on EBS must have a volume multiple
of 1 million of the base currency, which is therefore the natural
volume unit. This is, to our knowledge, the best data available from
EBS in terms of frequency (almost tick by tick) and, above all, has
the invaluable advantage of containing information about traded volumes.

\subsection{Treatment\label{sub:Treatment}}

\begin{figure}[H]
\centering{}\includegraphics[scale=0.3]{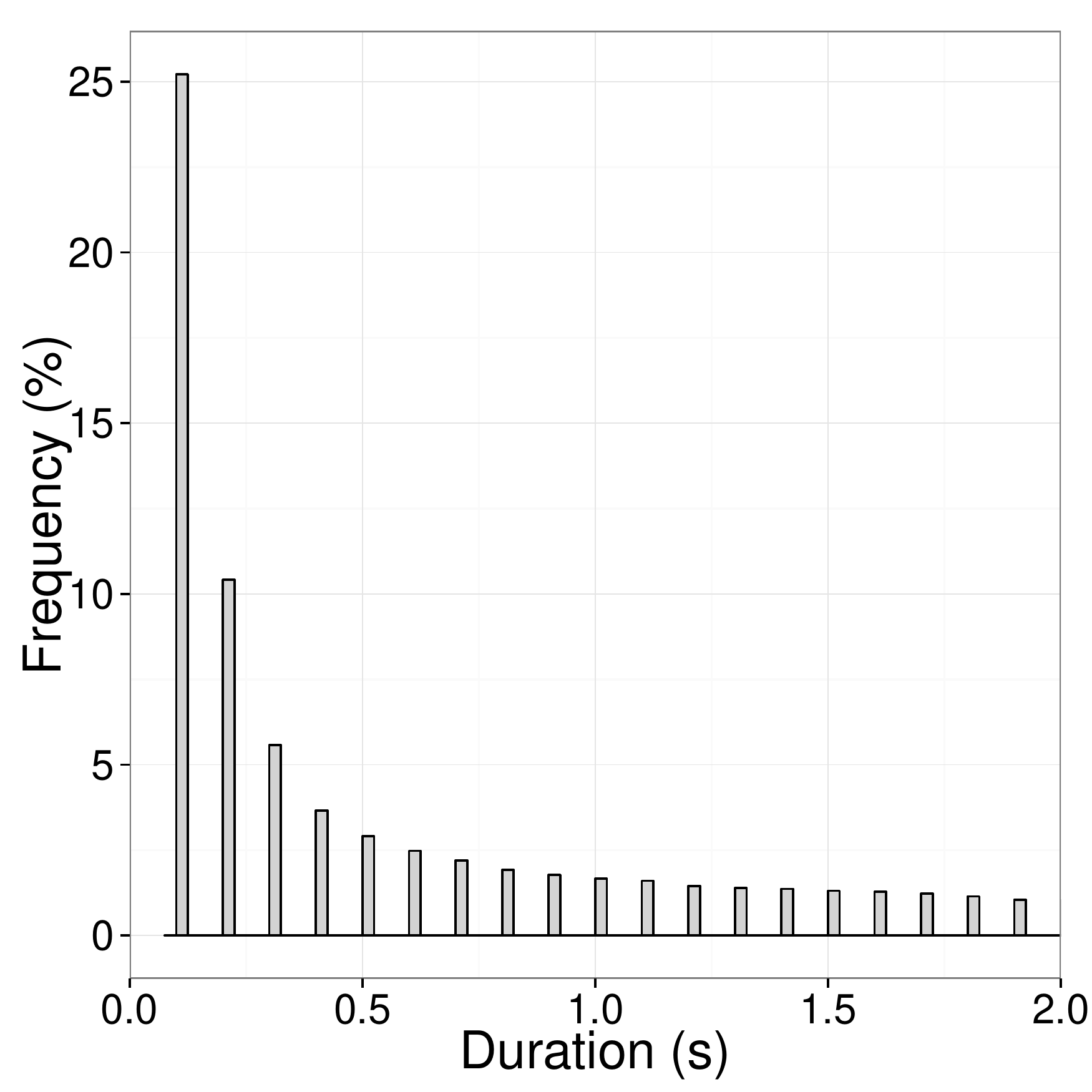}\includegraphics[scale=0.3]{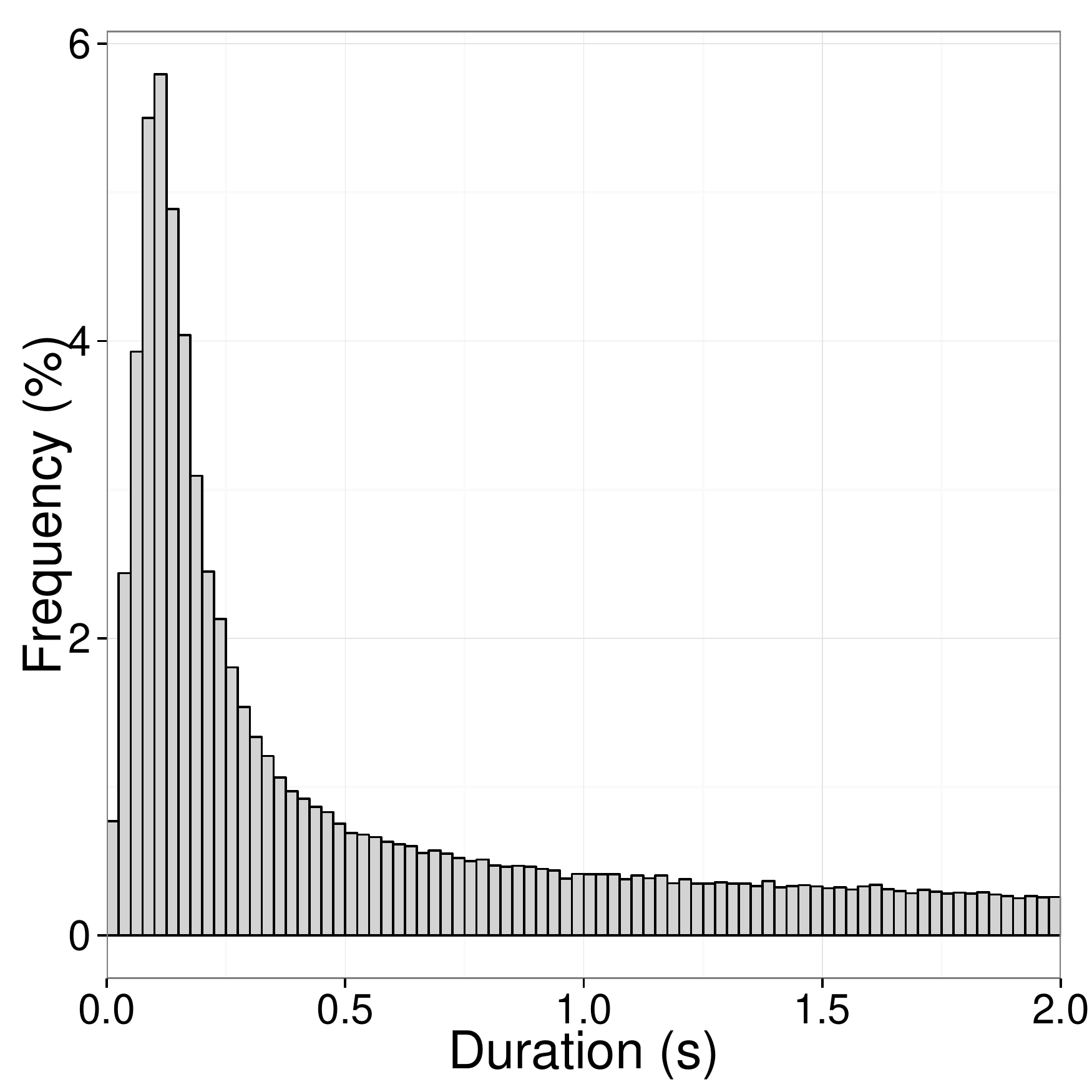}\includegraphics[scale=0.3]{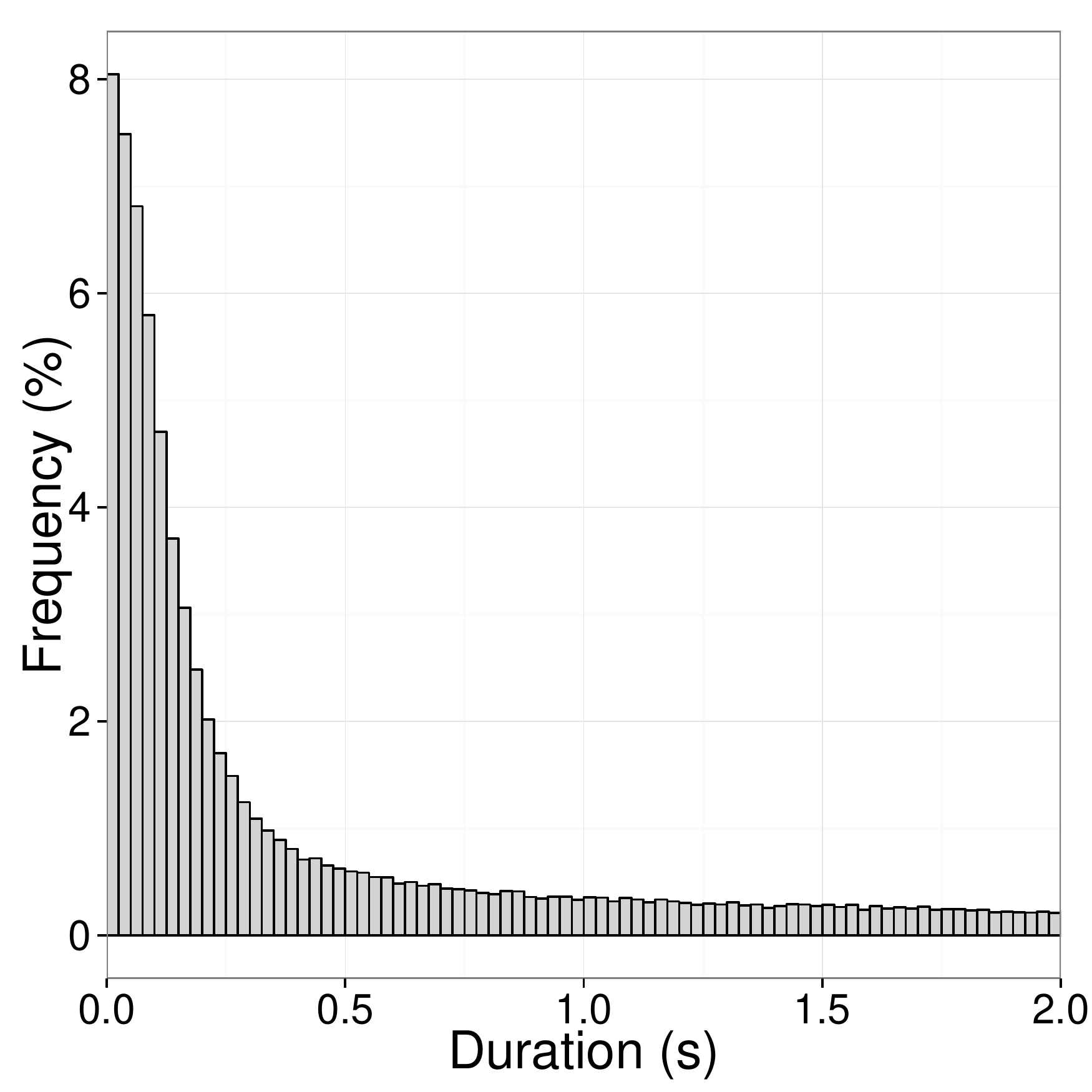}\protect\caption{\label{fig:durations}Duration distribution. Left: Raw times. Middle:
Randomized times. Right: Corrected times. Three months of data, restricted
to London active hours (9am-5pm).}
\end{figure}

The data must be filtered to improve the accuracy of fits. The coarse
time resolution introduces a spurious discretization of the duration
data, as illustrated in Fig. \ref{fig:durations} (left plot). To
overcome this issue, we added a time shift, uniformly distributed
between $0$ and $0.1$, to trade occurrence times (Fig. \ref{fig:durations},
middle plot). 

The number of transactions on one side during a time-slice can be
determined from the total signed volumes in $92\%$ of cases. Indeed,
when the total signed traded volume ($V_{total}$) is equal to the
reported trade volume ($V_{report}$), only one trade occurred and
the only uncertainty is about the exact time of the event. However,
when $V_{total}>V_{report}$, one knows that more than one trade occurred.
If $V_{total}-V_{report}=1$, exactly two trades occurred, one with
volume $V_{report}$ and one with volume 1; their respective event
time are randomly uniformly drawn during the time slice. Finally,
the case $V_{total}-V_{report}>1$ (about $8\%$ of the non-empty
time-slices) is ambiguous because the extra volume may come from more
than one trade and hence may be split in different ways. We tried
different schemes: not adding any trade, adding one trade, adding
a trade per extra million, adding a uniform random number of trades
between $1$ and $V_{total}-V_{report}$ and a self-consistent correction
that uses the most probable partition according to the distribution
of the volume of unambiguously determined trades. All of them give
similar estimated fitting parameters for all kernels. However, statistical
significance is best improved by adding one trade irrespective of
the kernel choice . We therefore apply this procedure in this paper;
as a consequence, all statistical results closely depend on this choice.
The distribution of resulting durations are plotted in Fig \ref{fig:durations}
(right plot). 

\begin{figure}[H]
\centering{}\includegraphics[scale=0.3]{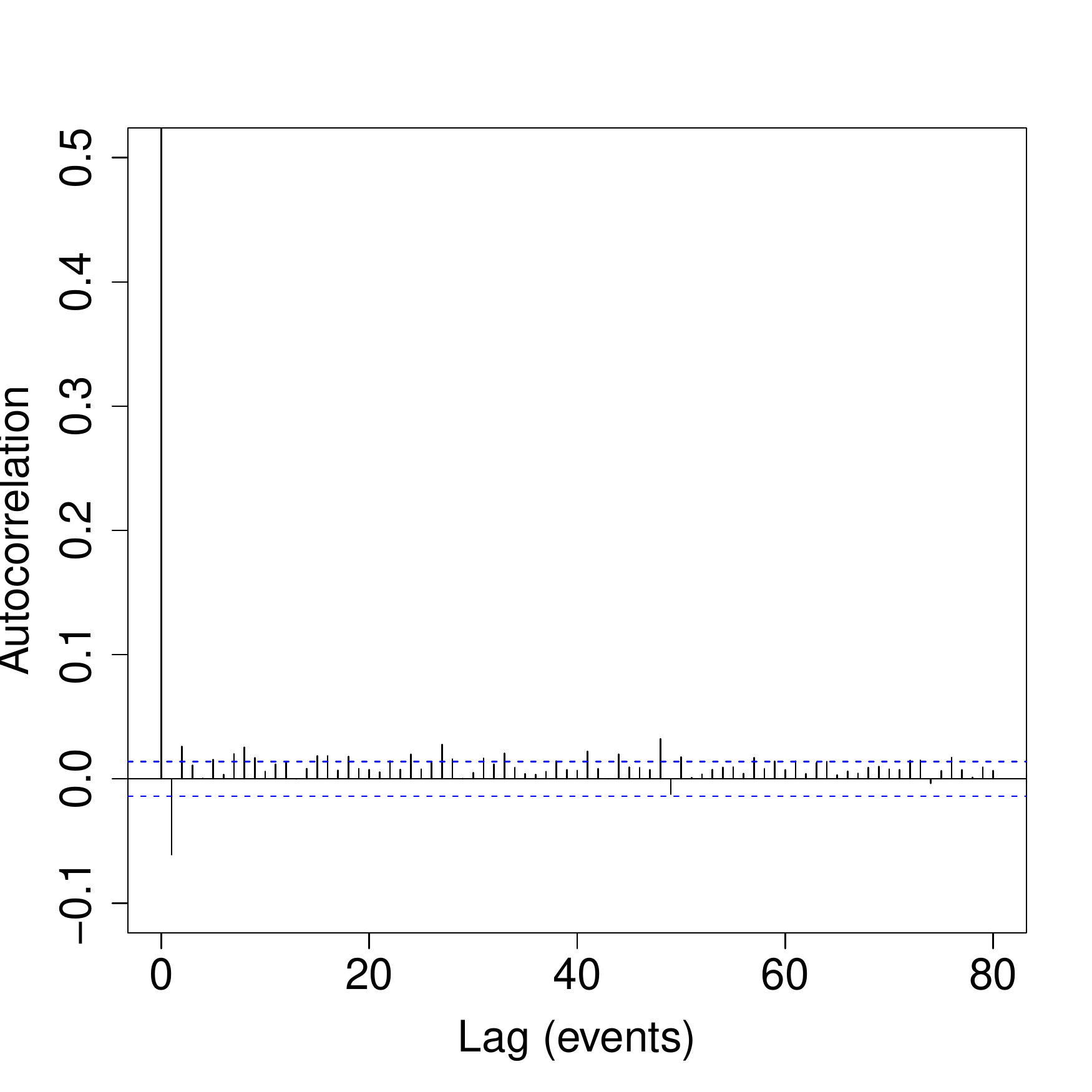}\includegraphics[scale=0.3]{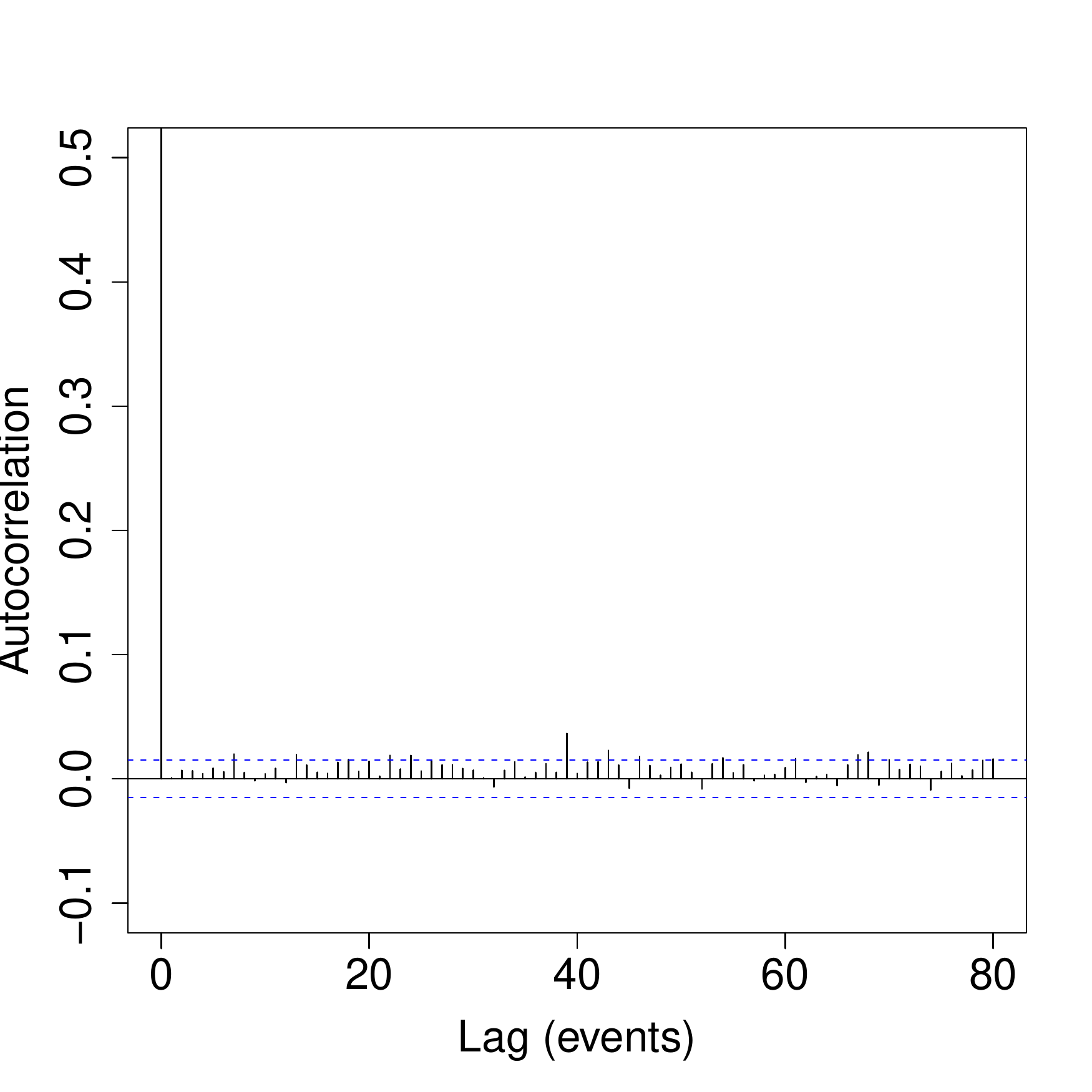}\protect\caption{\label{fig:acf}Time-adjusted durations autocorrelation function for
March 3rd 2012. Left: with correction. Right: without correction.}
\end{figure}

This simple correction procedure introduces a weak, short-term memory
effect. Figure \ref{fig:acf} (left) plots the linear auto-correlation
function of the sequence $\left\{ \theta_{i}\right\} $, for a particular
day (March 3rd 2012) (other days yield similar results). All the coefficients
are almost statistically equal to zero except at the first lag (which
is why we apply Ljung-Box test starting from the second lag). This
negative value is induced by the correction procedure (see Sec. \ref{sub:Treatment})
since the same measure performed in raw displays no memory at all
(Fig.~\ref{fig:acf} (right)). The auto-correlation of the$\left\{ (\theta_{i})^{2}\right\} $
series is however null with the correction procedure. This test therefore
shows that the time stamp correction procedure, without which no fit
ever passes a Kolmogorov-Smirnov test, is not entirely satisfactory
from this point of view. Nevertheless, the side effects are small
and most of the auto-correlation of the corrected timestamps is well
explained by a Hawkes model.

There may be other unwanted side effects caused by limited time resolution
and by the randomization of timestamps within a given interval. In
particular, one may wonder if limited time resolution introduces a
spurious small time scale in the fits. Appendix \ref{sec:appendix_sims}
reports extensive numerical simulations that assess the effect of
limited time resolution and time stamp shuffling and shows first that
this is not the case when time stamps are shuffled in an interval.
In addition, the smallest fitted time scale is influenced by the limited
time resolution, but to a limited extent.

\section{Results}

\subsection{Hourly fits\label{sub:Intra-day-fits}}

Hourly intervals are long enough to obtain reliable calibrations,
at least on active hours during which 1500 events take place on average.
In such short intervals, the endogenous activity $\mu_{t}$ in Eq.~\eqref{eq:HP_def}
can be approximated by a constant. We choose $m=2$ and $M=15$ for
the power-law types of kernel. At the hourly scale, the results are
fairly insensitive to changes in these parameters.

\subsubsection{Kernel comparisons}

Table \ref{tab:hourly} summarizes the results of the 8 types of kernels
for the three tests. The mono-exponential kernel $\phi_{1}$ is clearly
much worse than all the other specifications and we can safely rule
it out as a possible description of the data. Taking more than two
exponentials only marginally improves the fits of hourly activity.
QQ-plots (Fig. \ref{fig:gof}) illustrate the inadequacy of $\phi_{1}$
and show indeed that $\phi_{2}$ is a good kernel: for this time length,
two time scales are enough to describe a whole hour of the arrival
of FX trades. We judge the trade-off between log-likelihood and the
number of parameters with Akaike criterion, denoted by $AIC_{p}$
, Akaike weights $w_{i}$ of kernel $i$, and $N_{max}$, the number
of intervals in which kernel $i$ was the best. Both Akaike criteria
are averaged over all the intervals. In the end, both $w_{i}$ and
$N_{max}$ convey (almost) the same information because most of the
time only one kernel has a weight almost equal to one. Power-law types
of kernels also achieve good results, in particular $\phi_{15}^{{\scriptscriptstyle \text{PL}_{\text{x}}}}$,
but all indicate a larger endogeneity factor $n$ than kernels with
free exponentials. Akaike weights strongly suggest that $\phi_{2}$
is the best model at an hourly time scale. In addition we note that
the means and medians of the fitted parameters of $\phi_{n}$ ($n=1,2,3$)
kernels are very similar, while those of kernels that approximate
power laws are significantly different, which points to the fact that
this type of kernel is prone to fitting difficulties at an hourly
time scale. Finally, the free exponential of $\phi_{15}^{{\scriptscriptstyle \text{PL}_{\text{x}}}}$
has a timescale of \unit[0.06]{s}.

\begin{table}[H]
\begin{centering}
\begin{tabular}{|c|c|c|c|c|c|c|c|c|}
\hline 
 & $\phi_{1}$ & $\phi_{2}$ & $\phi_{3}$ & $\phi_{15}^{{\scriptscriptstyle \text{HBB}}}$ & $\phi_{15}^{{\scriptscriptstyle \text{PL}}}$ & $\phi_{30}^{{\scriptscriptstyle \text{HBB}}}$ & $\phi_{15}^{{\scriptscriptstyle \text{PL}}}$ & $\phi_{15}^{{\scriptscriptstyle \text{PL}_{\text{x}}}}$\tabularnewline
\hline 
\hline 
$\mu$ & $0.13$ & $0.08$ & $0.08$ & $0.07$ & $0.07$ & $0.07$ & $0.07$ & $0.06$\tabularnewline
\hline 
$n$ & $0.41$ & $0.64$ & $0.64$ & $0.67$ & $0.67$ & $0.77$ & $0.75$ & $0.72$\tabularnewline
\hline 
$\epsilon$ & NA & NA & NA & $0.23$ & $0.38$ & $0.26$ & $0.40$ & $0.28$\tabularnewline
\hline 
$pKS$ & $0.16$ & $0.69$ & $0.68$ & $0.56$ & $0.52$ & $0.56$ & $0.52$ & $0.56$\tabularnewline
\hline 
$pED$ & $0.03$ & $0.57$ & $0.55$ & $0.63$ & $0.60$ & $0.58$ & $0.57$ & $0.62$\tabularnewline
\hline 
$pLB$ & $0.11$ & $0.38$ & $0.38$ & $0.34$ & $0.31$ & $0.29$ & $0.28$ & $0.33$\tabularnewline
\hline 
$\log\mathcal{L}_{p}$ & $4022.9$ & $4069.5$ & $4069.9$ & $4055.6$ & $4062.9$ & $4045.8$ & $4060.3$ & $4064.2$\tabularnewline
\hline 
$AIC_{p}$ & $-8035.9$ & $-8122.7$ & $-8117.0$ & $-8098.2$ & $-8112.7$ & $-8078.5$ & $-8107.6$ & $-8105.8$\tabularnewline
\hline 
$w$ & $0.01$ & $0.55$ & $0.14$ & $0.05$ & $0.06$ & $0.03$ & $0.04$ & $0.11$\tabularnewline
\hline 
$N_{max}$ & $21$ & $692$ & $84$ & $65$ & $70$ & $21$ & $21$ & $116$\tabularnewline
\hline 
\end{tabular}
\par\end{centering}

\protect\caption{Comparison the ability of various kernels to fit Hawkes processes
on hourly time windows. $pKS$, $pED$ and $pLB$ are respectively
the Komogorov-Smirnov, Excess-Dispersion and Ljung-Box test average
p-values. $\log\mathcal{L}_{p}$ is the log likelihood per point for
the fit of each intervals, averaged over all intervals, and multiplied
by the average number of points per interval. Idem for the Akaike
information criterion $AIC_{p}$. The Akaike normalized weights $w[\phi]=\frac{1}{W}\exp\left(-\frac{AIC[\phi]-AIC_{min}}{2}\right)$,
are the probabilites that kernel $\phi$ is the best according to
Kullback\textendash Leibler discrepancy \citep{Wagenmakers2004}.
$N_{max}[\phi]$ is the number of intervals in which the Akaike weight
of kernel $\phi$ is the largest one. Values averaged over the fits
on 1090 non-overlapping windows with more than 200 trades. \label{tab:hourly}}
\end{table}

\subsubsection{Detailed results for $\phi_{2}$}

Given its simplicity and good performance, it is interesting to look
further into the results for the double exponential case. We note
that \citet{Rambaldi2014} also suggest that this kernel is a good
candidate for the modeling of mid-quotes changes in EBS data (without
signed volumes). We characterize each hourly time-window by averaging
the fits over three months.

\begin{figure}[H]
\centering{}\includegraphics[scale=0.276]{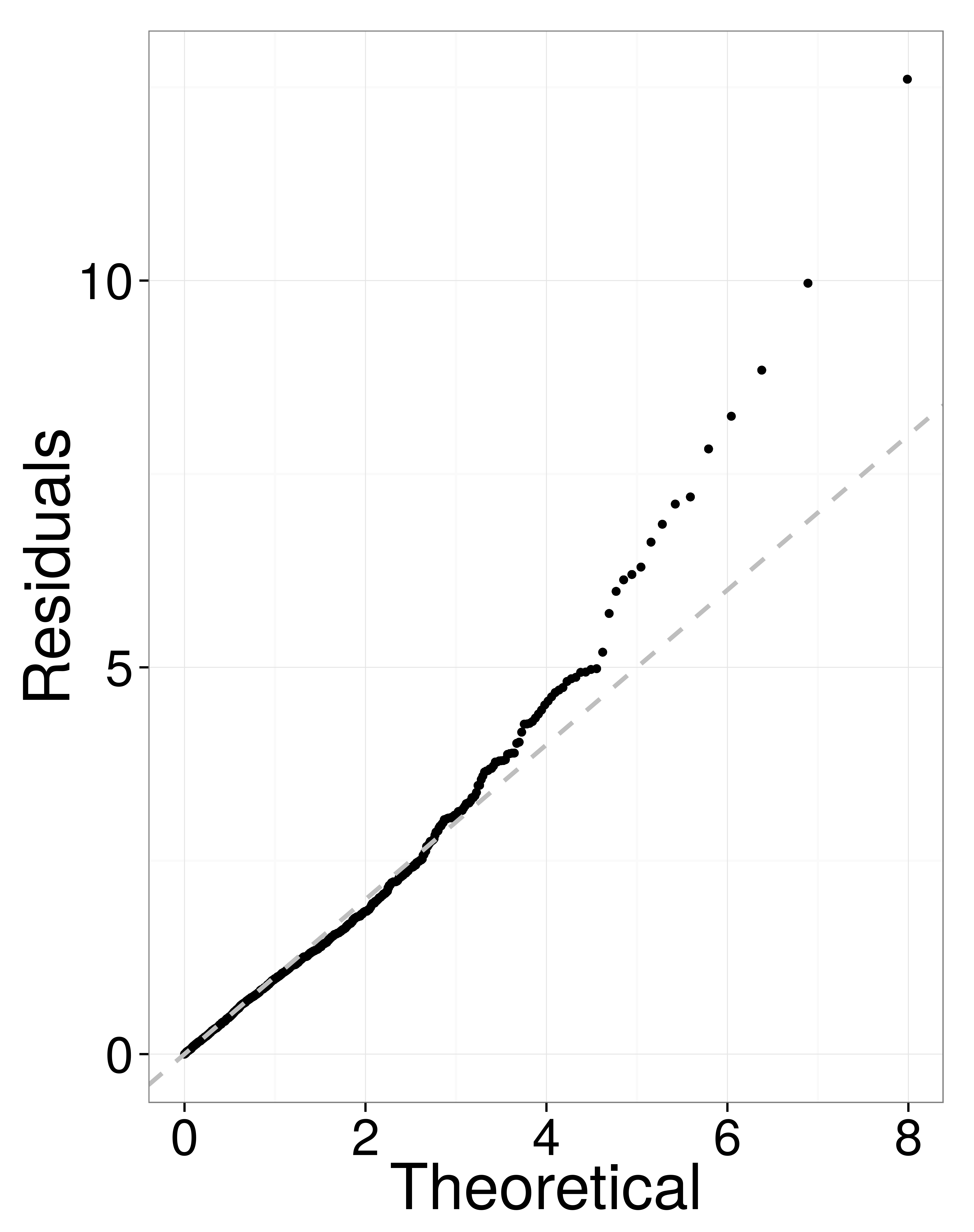}\includegraphics[scale=0.3]{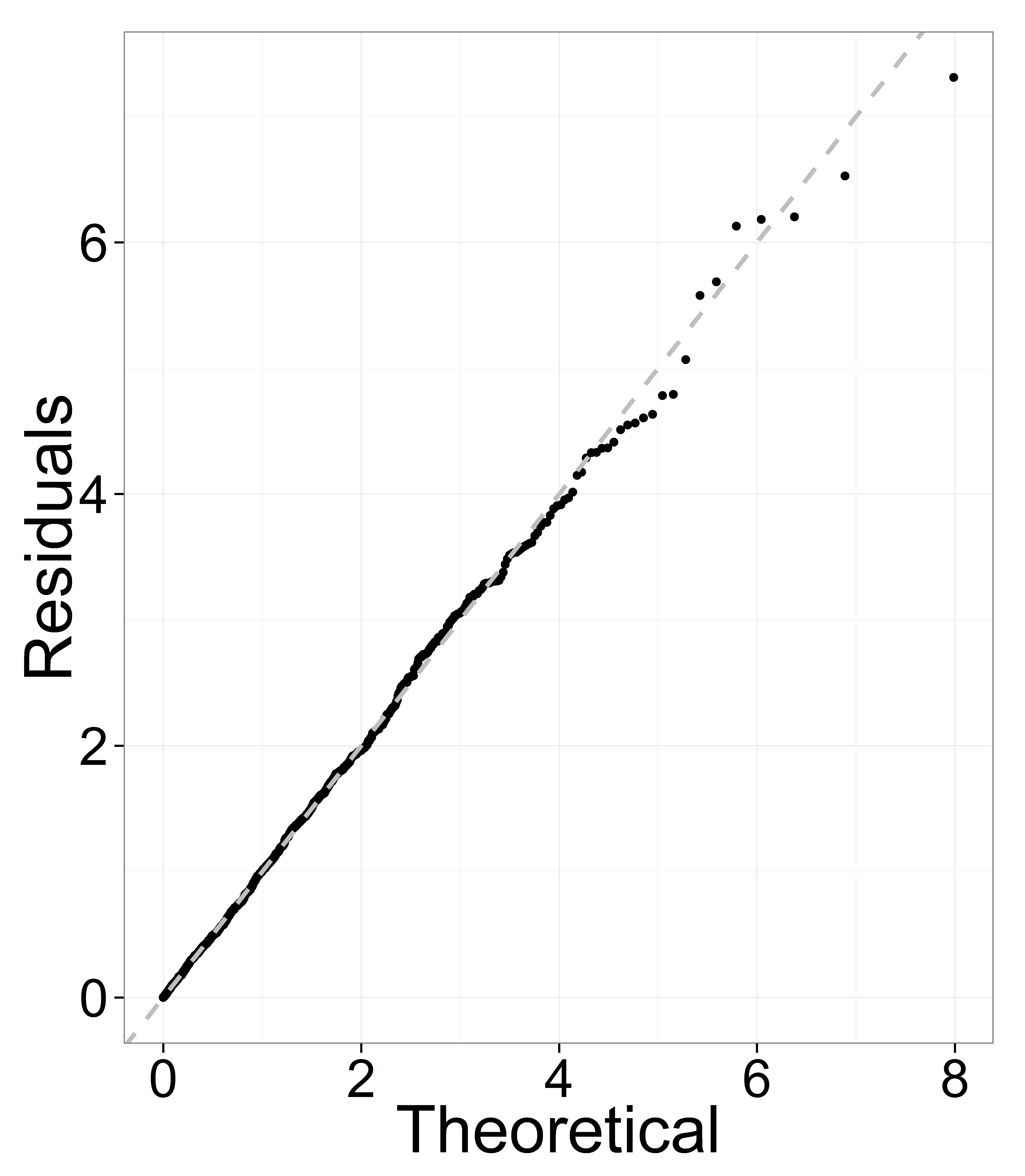}\includegraphics[scale=0.275]{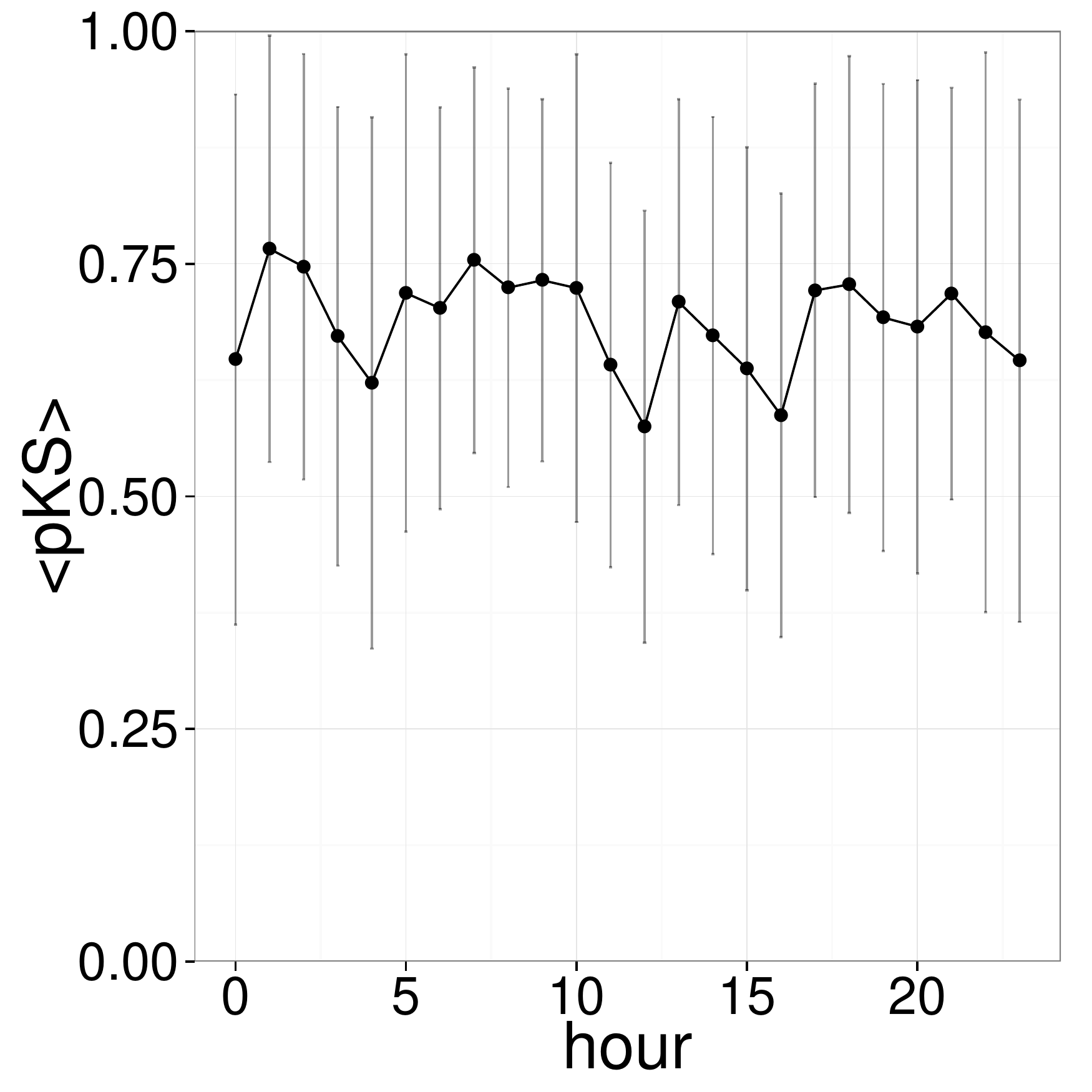}\protect\caption{\label{fig:gof}Goodness of Fit tests under the null hypothesis of
exponentially distributed time-deformed durations. Left: A typical
QQ-plot (February 1st 2012, 3-4pm) for $\phi_{1}$. Middle: Same for
$\phi_{2}$. Right: Kolmogorov-Smirnov test average p-value. Error
bars set at two standard deviations.}
\end{figure}

First, let us have a look at goodness of fits results. Fig. \ref{fig:gof}
(left plot) reports the quantiles of $\left\{ \theta_{i}\right\} $
for a particular day and hourly window against the exponential theoretical
quantiles. The fit is visually very satisfactory. Other time windows
of all days yield similar results. Fig. \ref{fig:gof} (right plot)
demonstrates that all hours of the day pass Kolmogorov-Smirnov test
by a large margin.

\begin{figure}[H]
\centering{}\includegraphics[scale=0.3]{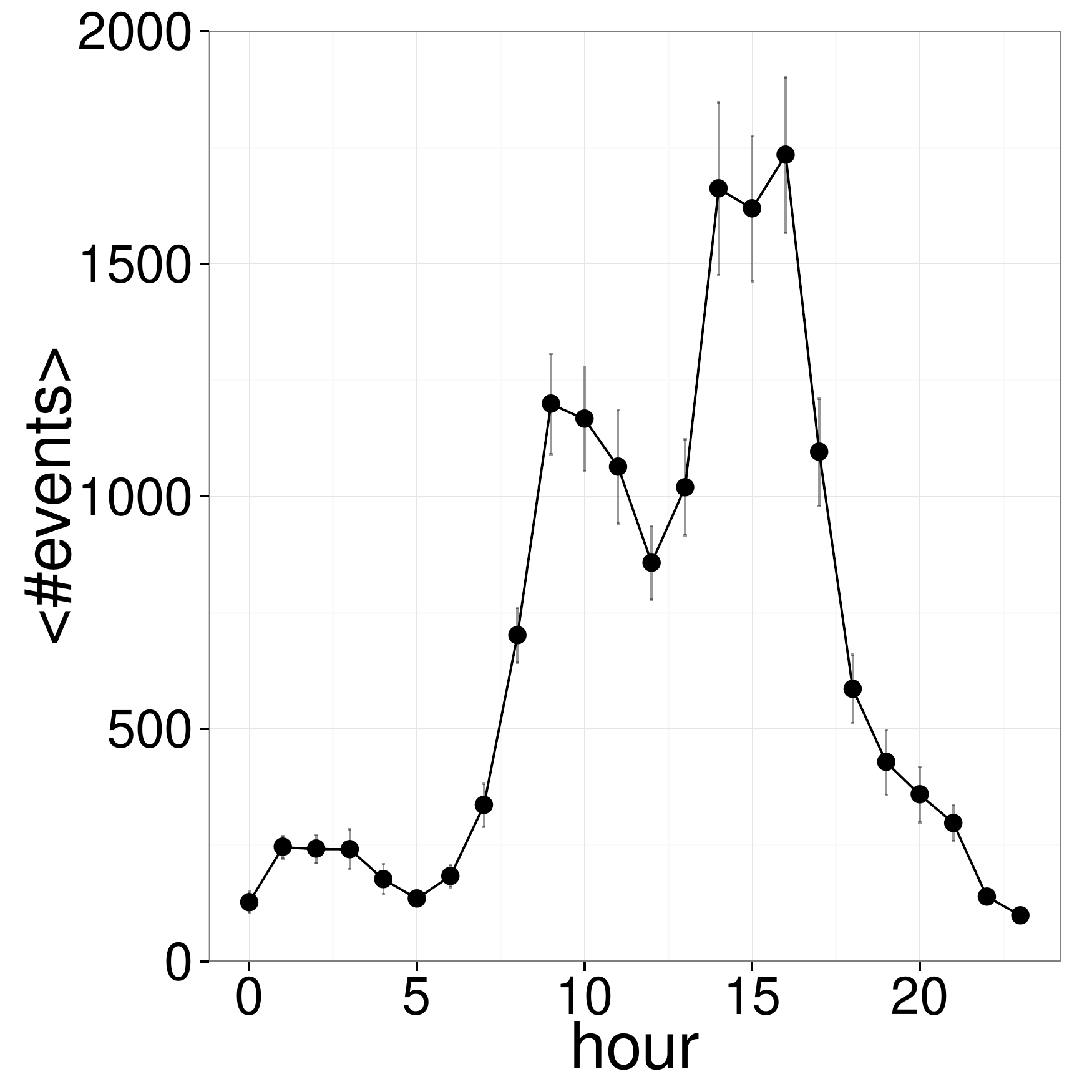}\includegraphics[scale=0.3]{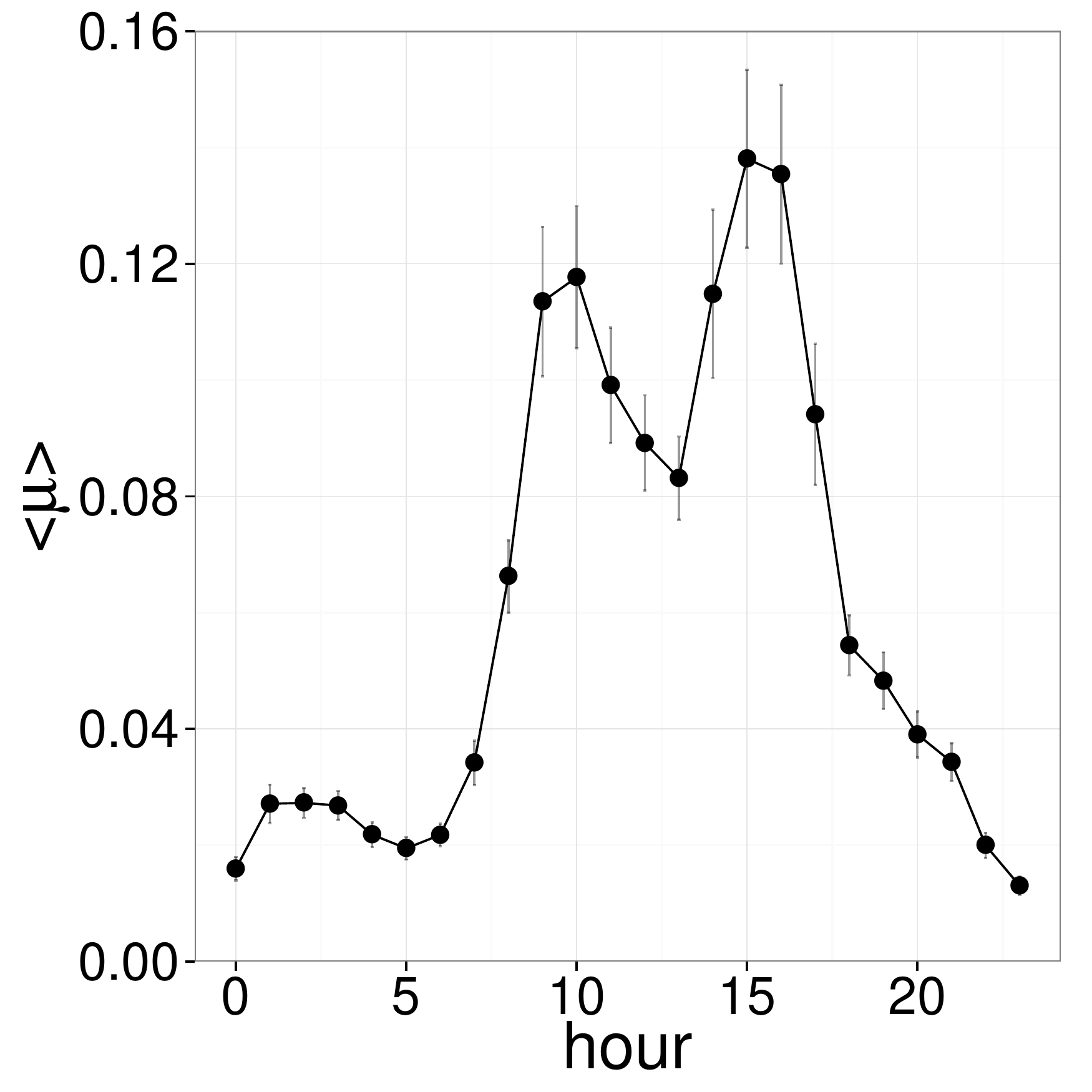}\protect\caption{\label{fig:mu}Left: Average number of trades (left) and average baseline
intensity (right) throughout the day. Error bars set at two standard
deviations.}
\end{figure}

In Fig. \ref{fig:mu} (left), the number of trades displays the well-known
intraday pattern of activity in the FX market \citep{Dacorogna1993,Ito2006}.
The average fitted exogenous part $\left\langle \mu\right\rangle $
perfectly reproduces this activity pattern (Fig. \ref{fig:mu}, right
plot).

\begin{figure}[H]
\centering{}\includegraphics[scale=0.3]{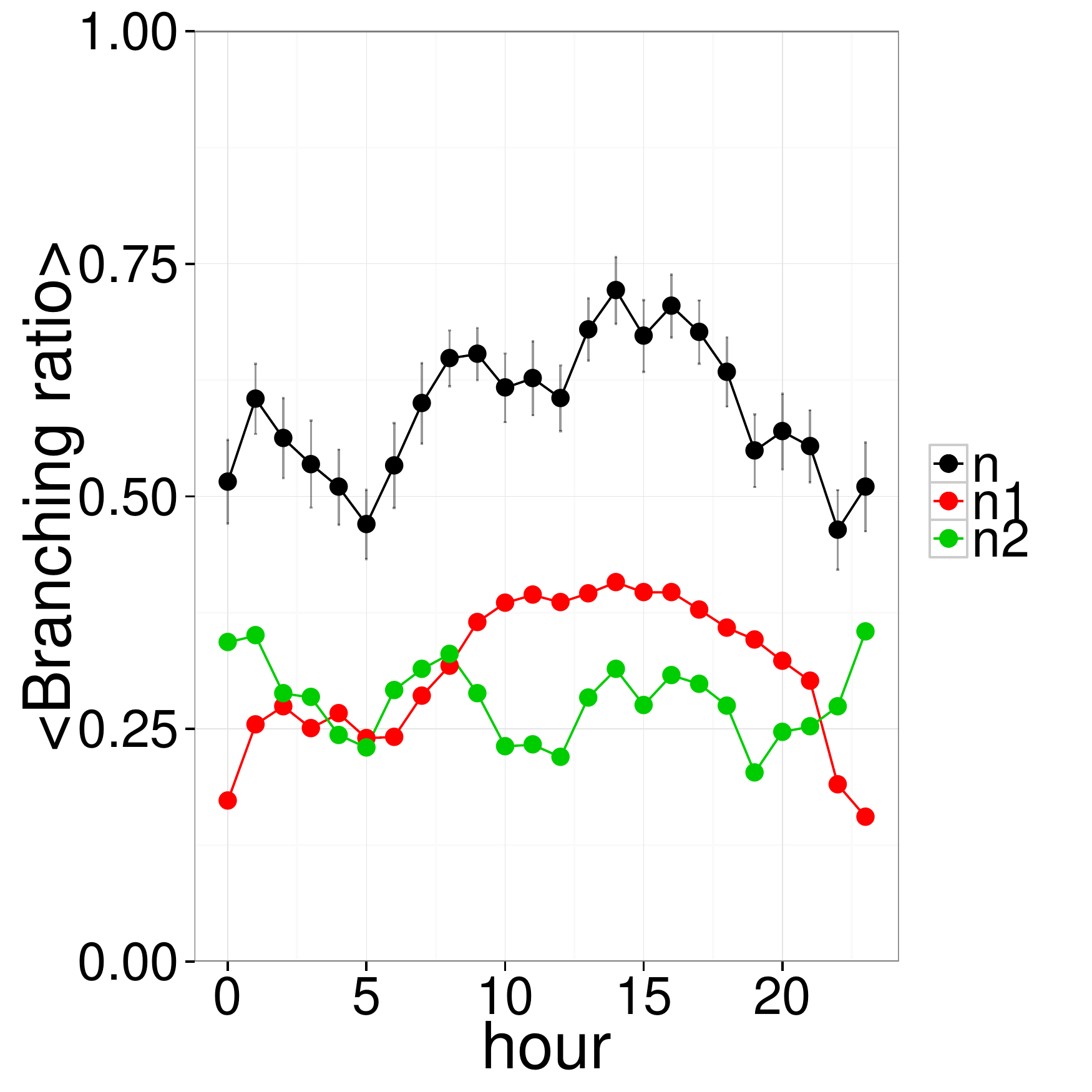}\includegraphics[scale=0.3]{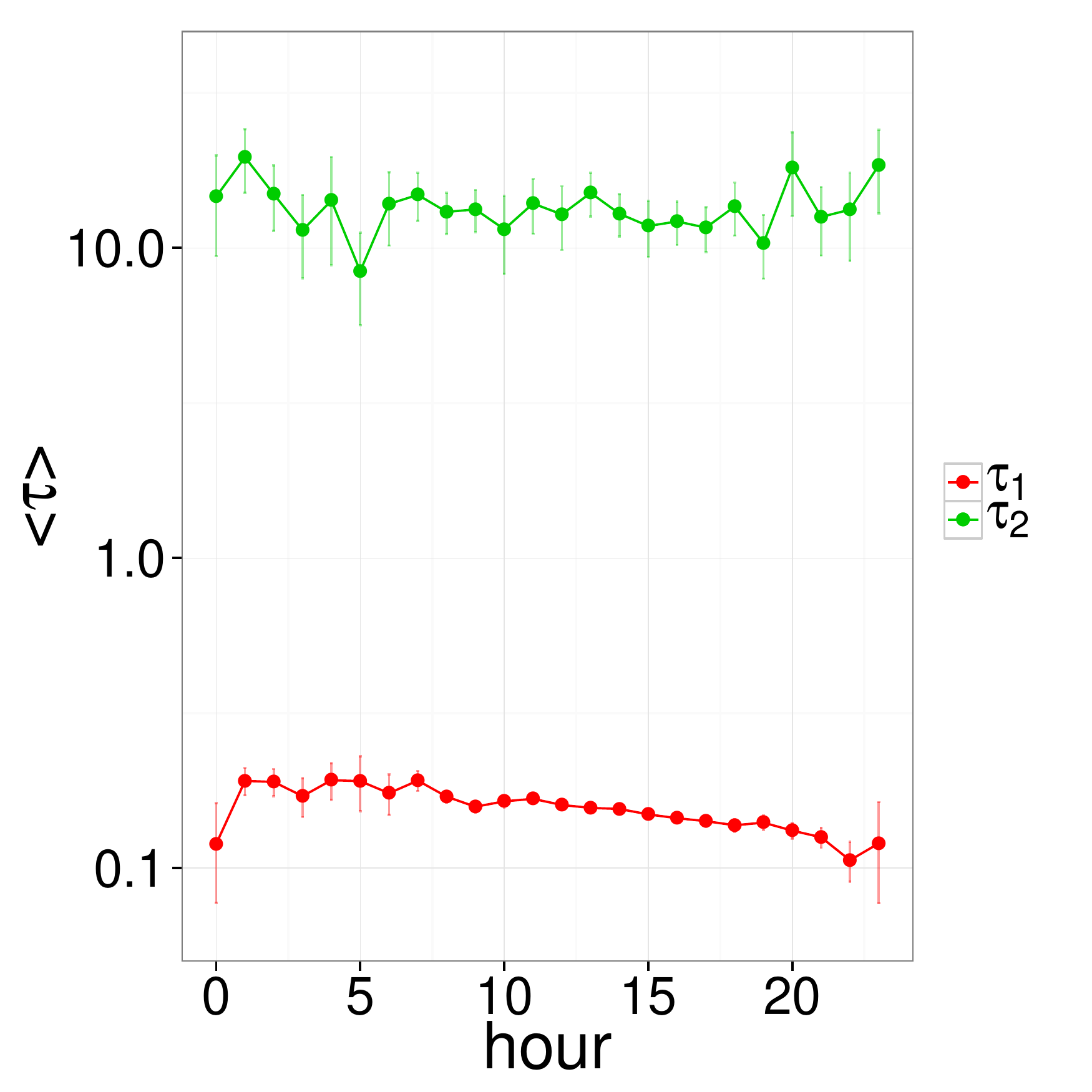}\protect\caption{\label{fig:Branching-ratio}Average branching ratio throughout the
day (left); black symbols: total ratio; green symbols: branching ratio
of the largest time-scale; blue symbols: branching ratio of the smallest
time-scale. Average associated times-scales on the right. Error bars
set at two standard deviations.}
\end{figure}

Remarkably, the endogeneity level $n$ is relatively stable (within
statistical uncertainty) for all hours (Fig. \ref{fig:Branching-ratio})
given the fact that the typical trading activity is 10 times smaller
at nights (Fig. \ref{fig:mu}). This is particularly striking for
the endogeneity associated to largest time scale, $n_{2}$. Endogeneity
associated with the smallest time scale, $n_{1}$, follows, albeit
with a much smaller relative change, the daily average activity, except
for the lunch time lull, which comes from the largest time scale.
This suggests that while automated algorithmic trading takes no pause,
human traders do have a break. In turn, this means that at this scale,
most of the endogeneity at the smallest time scale comes from algorithmic
trading, and that a sizable part of the endogeneity at longer times
scales is caused by human trading.

\subsection{Whole-day fits\label{sub:Daily-fit}}

The relative stability of the branching ratio and the high p-values
of e.g. KS tests encourages us to fit longer time windows. As we will
see, this is possible for a full day at a time. In this case, $\mu$
cannot be considered constant anymore (see Fig. \ref{fig:mu}). As
suggested by \citet{Bacry2014a}, a time-of-the-day dependent background
intensity is a good way to account for the intraday variation of activity.
This method has the advantage of not mixing data from other days like
classic detrending methods do. We thus approximate, for each day,
$\mu_{t}$ by a piecewise linear function with knots at 0~am (when
the series begin), 5~am, 9~am, 12~pm, 4~pm and at the end of the
series. The $6$ knots values are additional fitting parameters.

\subsubsection{Kernel comparison}

The results are synthesized in table \ref{tab:table_whole}.

\begin{table}[H]
\begin{centering}
\begin{tabular}{|c|c|c|c|c|c|c|c|c|c|c|c|}
\hline 
Kernel & $\phi_{1}$ & $\phi_{2}$ & $\phi_{3}$ & $\phi_{4}$ & $\phi_{15}^{{\scriptscriptstyle \text{HBB}}}$ & $\phi_{15}^{{\scriptscriptstyle \text{PL}}}$ & $\phi_{15}^{{\scriptscriptstyle \text{HBB}}}/\mu_{cst}$ & $\phi_{15}^{{\scriptscriptstyle \text{PL}}}/\mu_{cst}$ & $\phi_{30}^{{\scriptscriptstyle \text{HBB}}}$ & $\phi_{30}^{{\scriptscriptstyle \text{PL}}}$ & $\phi_{15}^{{\scriptscriptstyle \text{PL}_{\text{x}}}}$\tabularnewline
\hline 
\hline 
$n$ & $0.48$ & $0.79$ & $0.83$ & $0.85$ & $0.81$ & $0.83$ & $0.92$ & $0.93$ & $0.98$ & $0.97$ & $0.88$\tabularnewline
\hline 
$pKS$ & $7e-13$ & $0.09$ & $0.13$ & $0.16$ & $4\e{-6}$ & $2\e{-7}$ & $6\e{-9}$ & $6\e{-10}$ & $6\e{-4}$ & $4e-6$ & $0.04$\tabularnewline
\hline 
$pED$ & $0$ & $0.10$ & $0.31$ & $0.45$ & $0.61$ & $0.51$ & $0.6$ & $0.54$ & $0.52$ & $0.49$ & $0.66$\tabularnewline
\hline 
$pLB$ & $0$ & $0.058$ & $0.056$ & $0.012$ & $9e-5$ & $0.001$ & $0.017$ & $0.026$ & $1\e{-4}$ & $2e-4$ & $0.006$\tabularnewline
\hline 
$\log\mathcal{L}_{p}$ & $60271.0$ & $61559.3$ & $61596.2$ & $61575.5$ & $61279.6$ & $61340.2$ & $61181.5$ & $61271.1$ & $61286.3$ & $61340.6$ & $61468$\tabularnewline
\hline 
$AIC_{p}$ & $-120525.4$ & $-123097.7$ & $-123167.4$ & $-123121.8$ & $-122540.4$ & $-122661.7$ & $-122354.7$ & $-122534.0$ & $-122553.9$ & $-122662.5$ & $-122912.0$\tabularnewline
\hline 
$\epsilon$ & NA & NA & NA & NA & $0.090$ & $0.115$ & $0.027$ & $0.057$ & $0.13$ & $0.14$ & $0.08$\tabularnewline
\hline 
$w$ & $0$ & $0.13$ & $0.38$ & $0.24$ & $0$ & $0$ & $0$ & $0$ & $0$ & $0$ & $0.24$\tabularnewline
\hline 
$N_{max}$ & $0$ & $9$ & $22$ & $14$ & $0$ & $0$ & $0$ & $0$ & $0$ & $0$ & $14$\tabularnewline
\hline 
\end{tabular}
\par\end{centering}

\protect\caption{Kernel comparison. Full day fits. 59 points.\label{tab:table_whole}}
\end{table}

\begin{figure}[H]
\begin{centering}
\includegraphics[scale=0.3]{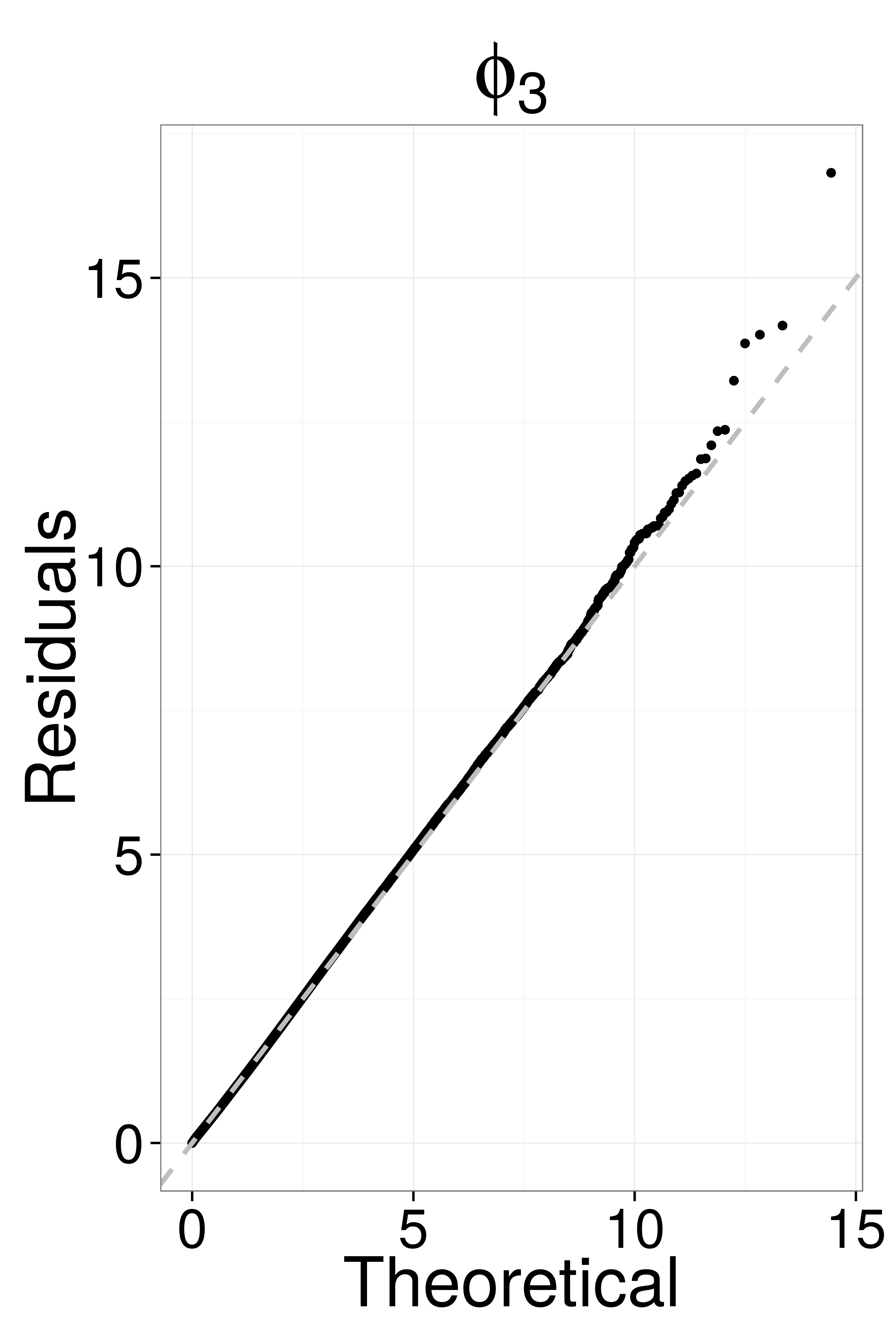}\includegraphics[scale=0.3]{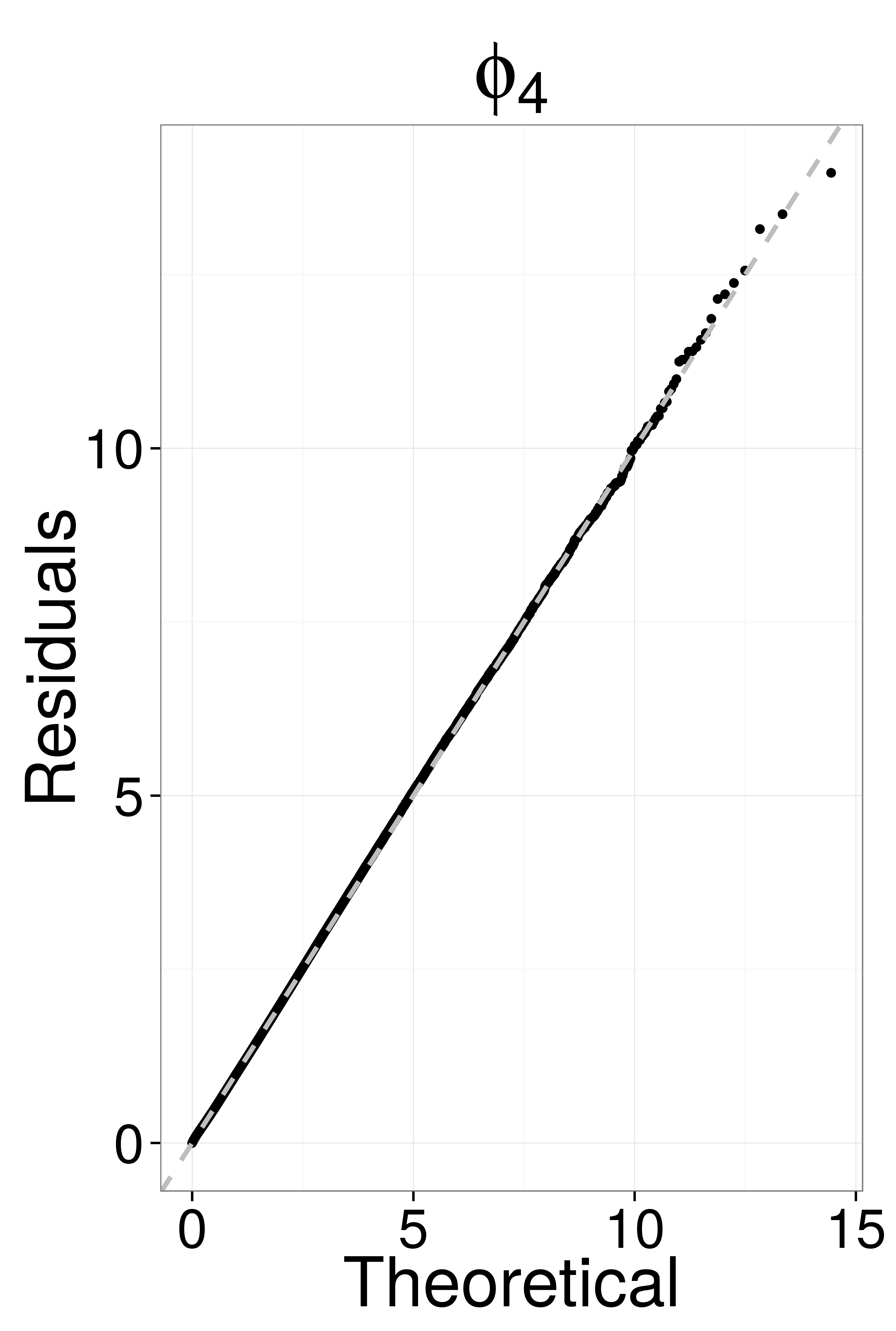}
\par\end{centering}

\begin{centering}
\includegraphics[scale=0.3]{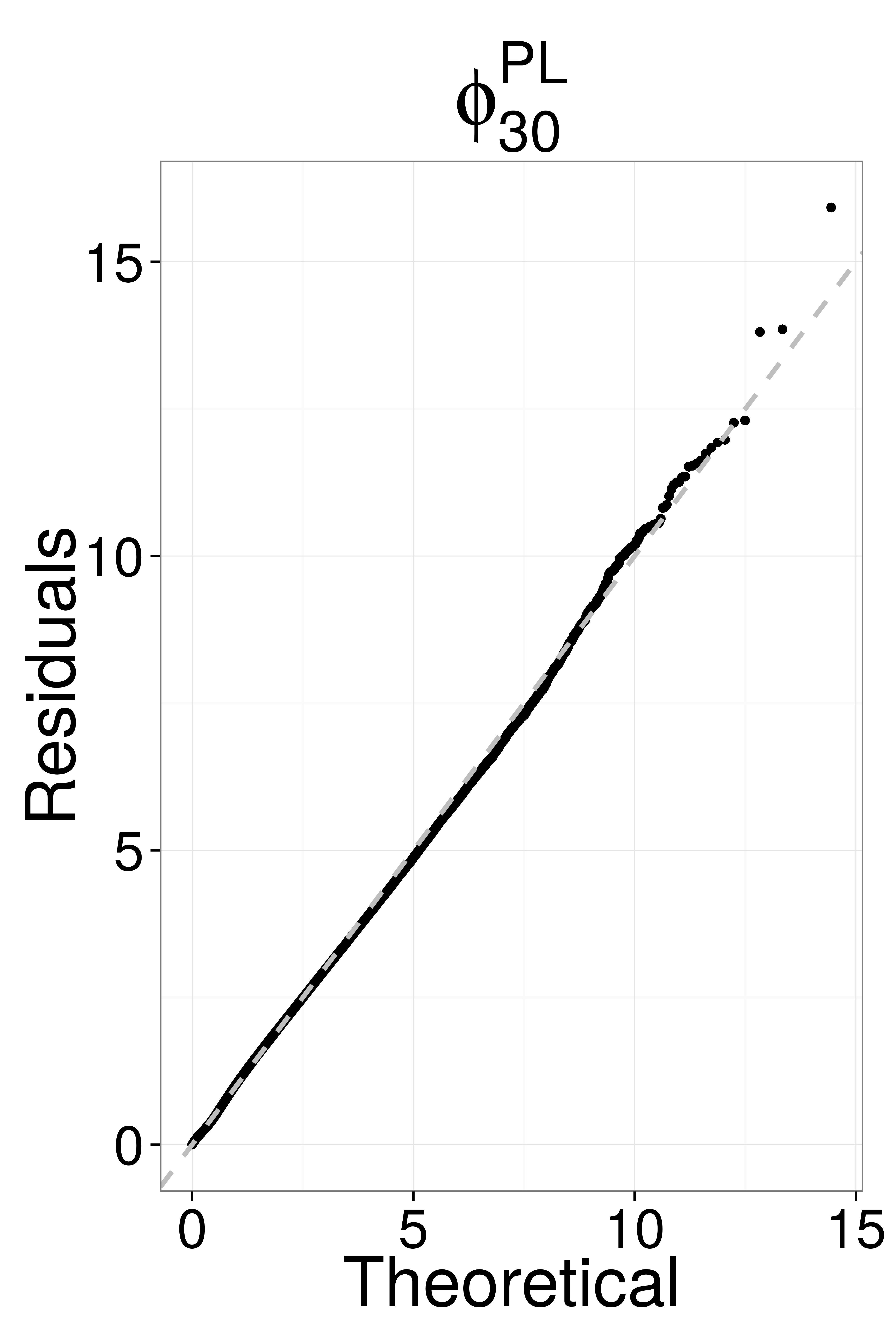}\includegraphics[scale=0.3]{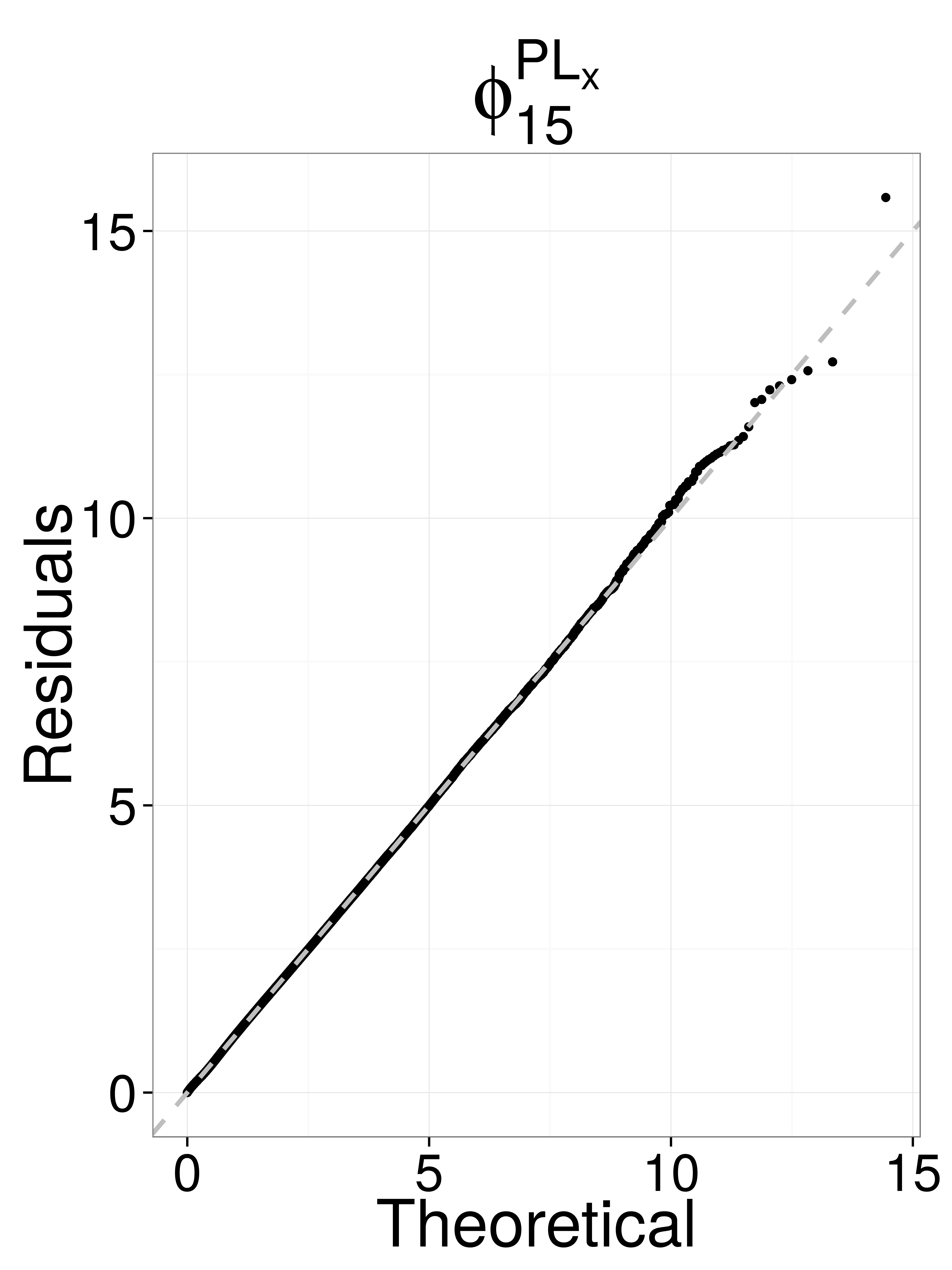}
\par\end{centering}

\centering{}\protect\caption{QQ-plot of the residuals merged from all intervals (one-day fits).
\label{fig:QQplot_merge}}
\end{figure}

Only $\phi_{2}$ and $\phi_{3}$ pass the Ljung-Box test. This time
$\phi_{3}$ is the favored model according to the Akaike weights and
performs well with respect to the three tests. We note that $\phi_{15}^{{\scriptscriptstyle \text{PL}_{\text{x}}}}$,
whose free exponential has a timescale equal to \unit[0.11]{s}, is
also a strong contender. We can gain a global insight across days
from QQplots. Indeed, under the null hypothesis, the residuals possess
the same distribution independently of the considered day. We therefore
merge all the residuals from all the daily fits and construct the
QQplot against the exponential distribution. Fig.~\ref{fig:QQplot_merge}
reports the performance of four families of kernel and bring a visual
confirmation of the results in Table \ref{tab:table_whole}. In addition,
it allows one to understand where each kernel performs best and worst.
For example, $\phi_{30}^{{\scriptscriptstyle \text{PL}}}$ is better
in the extreme tails than in the bulk of the distribution. One also
sees the problems of $\phi_{3}$ in this region, solved by adding
a fourth exponential (see $\phi_{4}$).

\subsubsection{Detailed results for $\phi_{3}$}

\begin{figure}[H]
\begin{centering}
\includegraphics[scale=0.35]{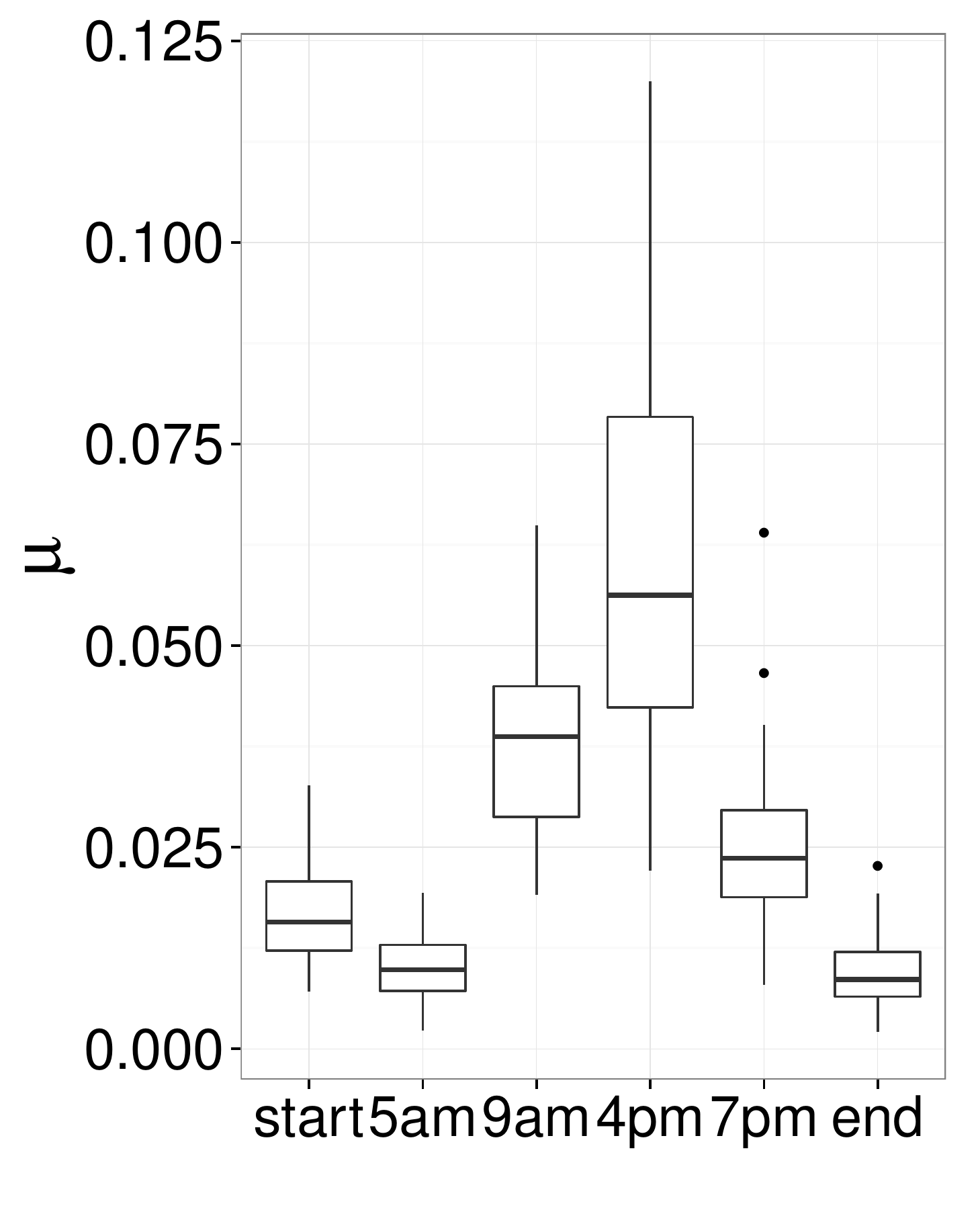}
\par\end{centering}

\protect\caption{\label{fig:Baseline-intensity-knots.}Tukey boxplot of baseline intensity
knots values.}
\end{figure}

Let us investigate in details the fits of $\phi_{3}$, the overall
best kernel for whole days. We also show some results for $\phi_{2}$
for sake of comparison. The background intensity fitted values are
summarized in Fig. \ref{fig:Baseline-intensity-knots.} and are in
line with the average intraday activity pattern.

\begin{figure}[H]
\begin{centering}
\includegraphics[scale=0.3]{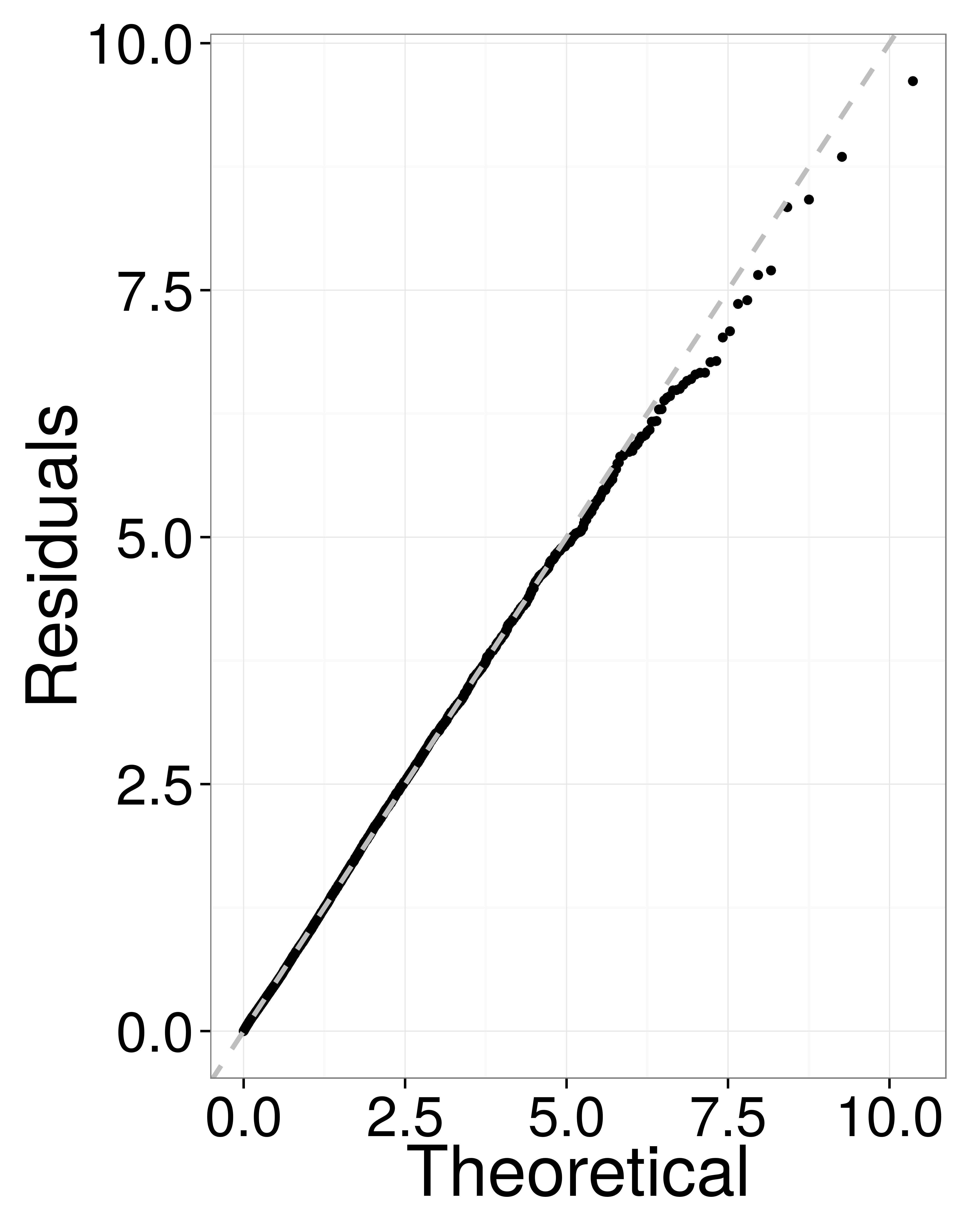}\includegraphics[scale=0.3]{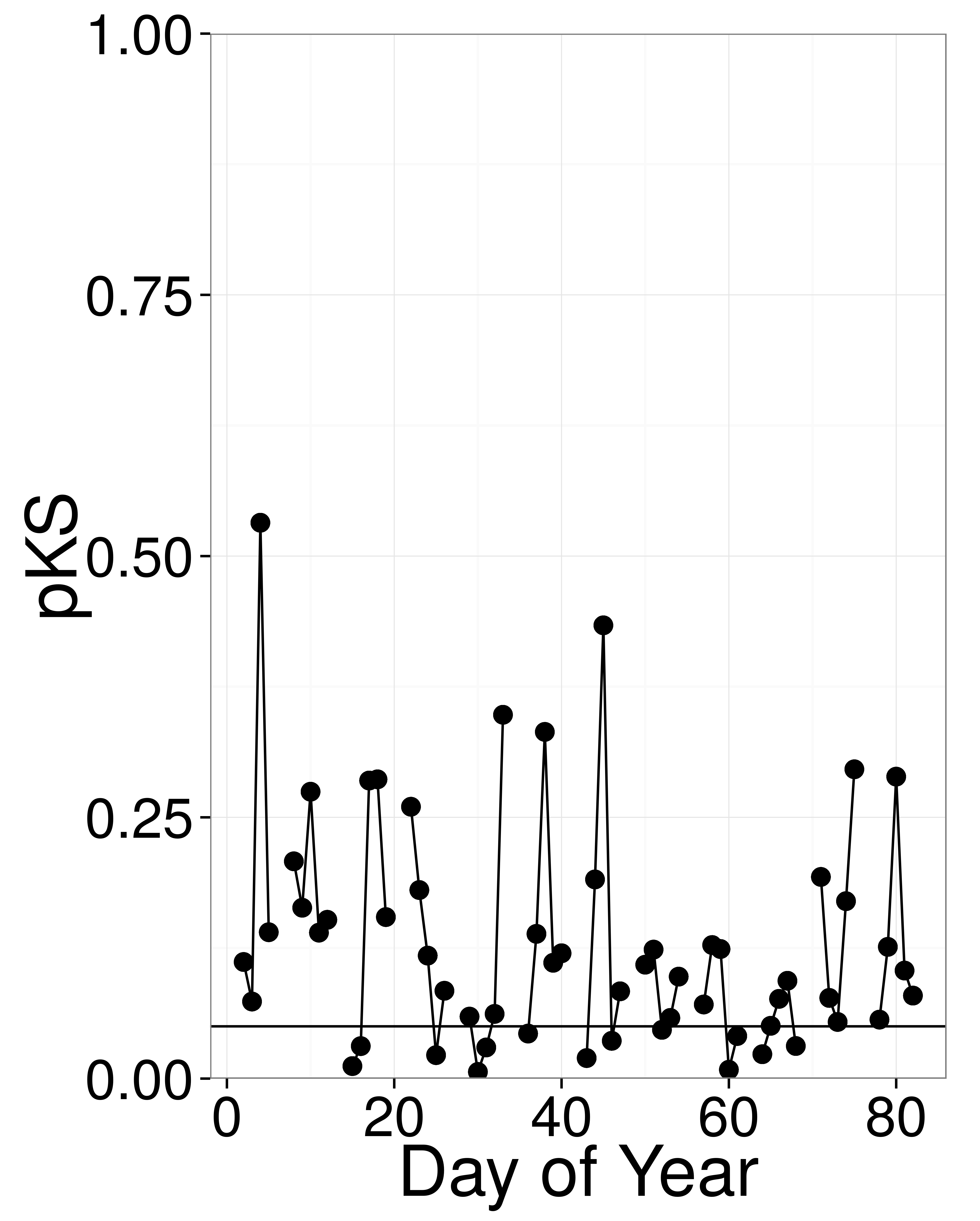}
\par\end{centering}

\centering{}\protect\caption{\label{fig:res}Goodness of Fit tests under the null hypothesis of
exponentially distributed time-deformed durations. Left: A typical
QQ-plot (March 3rd 2012). Right: Kolmogorov-Smirnov test p-value.
The continuous line is the 0.05 significance level.}
\end{figure}

Figure \ref{fig:res} reports the Kolmogorov-Smirnov p-value for each
fitted day. Again, the null hypothesis of exponentially distributed
$\left\{ \theta_{i}\right\} $, i.e., good fits, cannot be rejected.
Fits are however less impressively significant that those of hourly
fits case because of additional non-stationarities. On this plot and
on all the remaining plots of the section, line breaks correspond
to weekends. The QQ-plot (left plot of Fig.~\ref{fig:res}) visually
confirms the accuracy of the fit.

\begin{figure}[H]
\centering{}\includegraphics[scale=0.35]{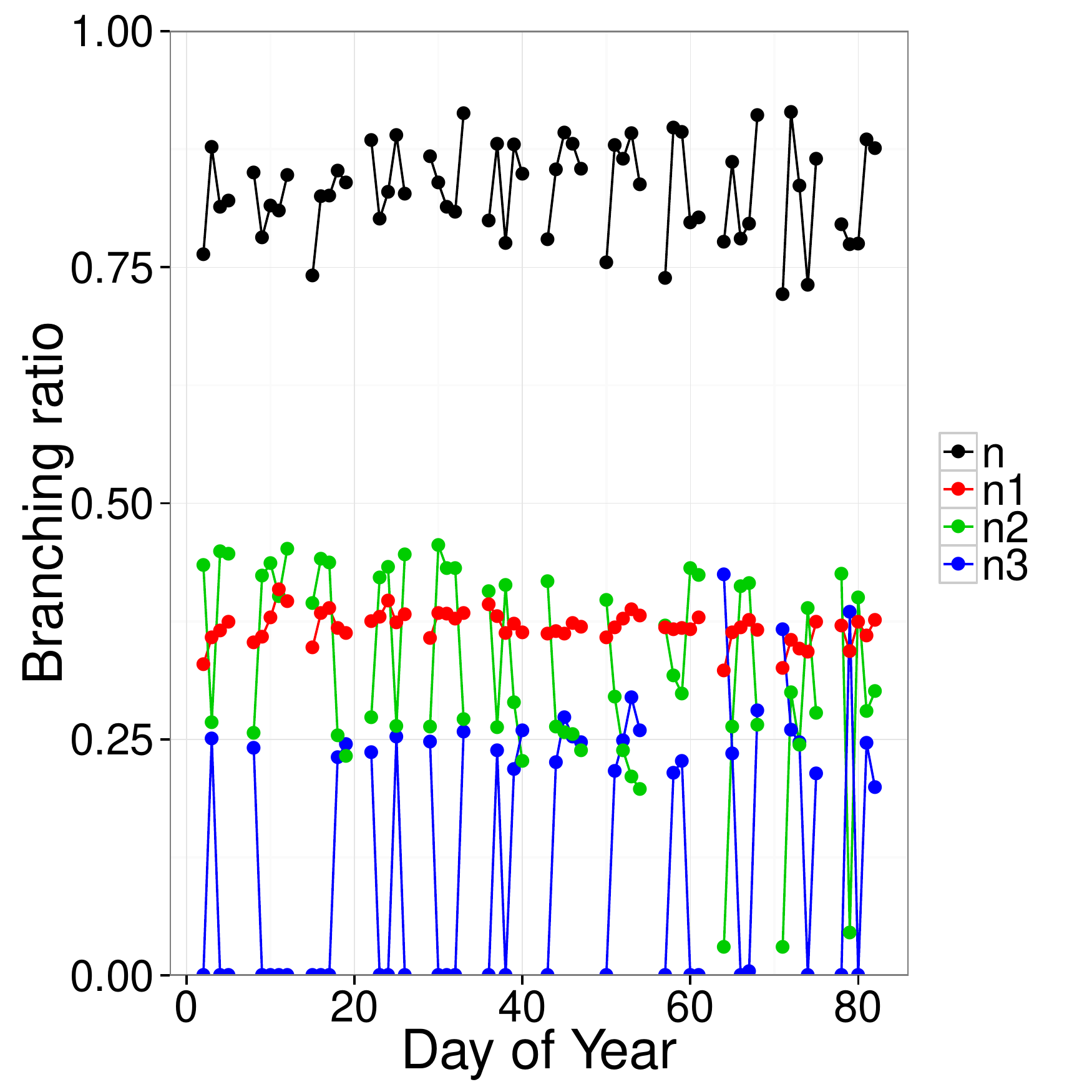}\includegraphics[scale=0.35]{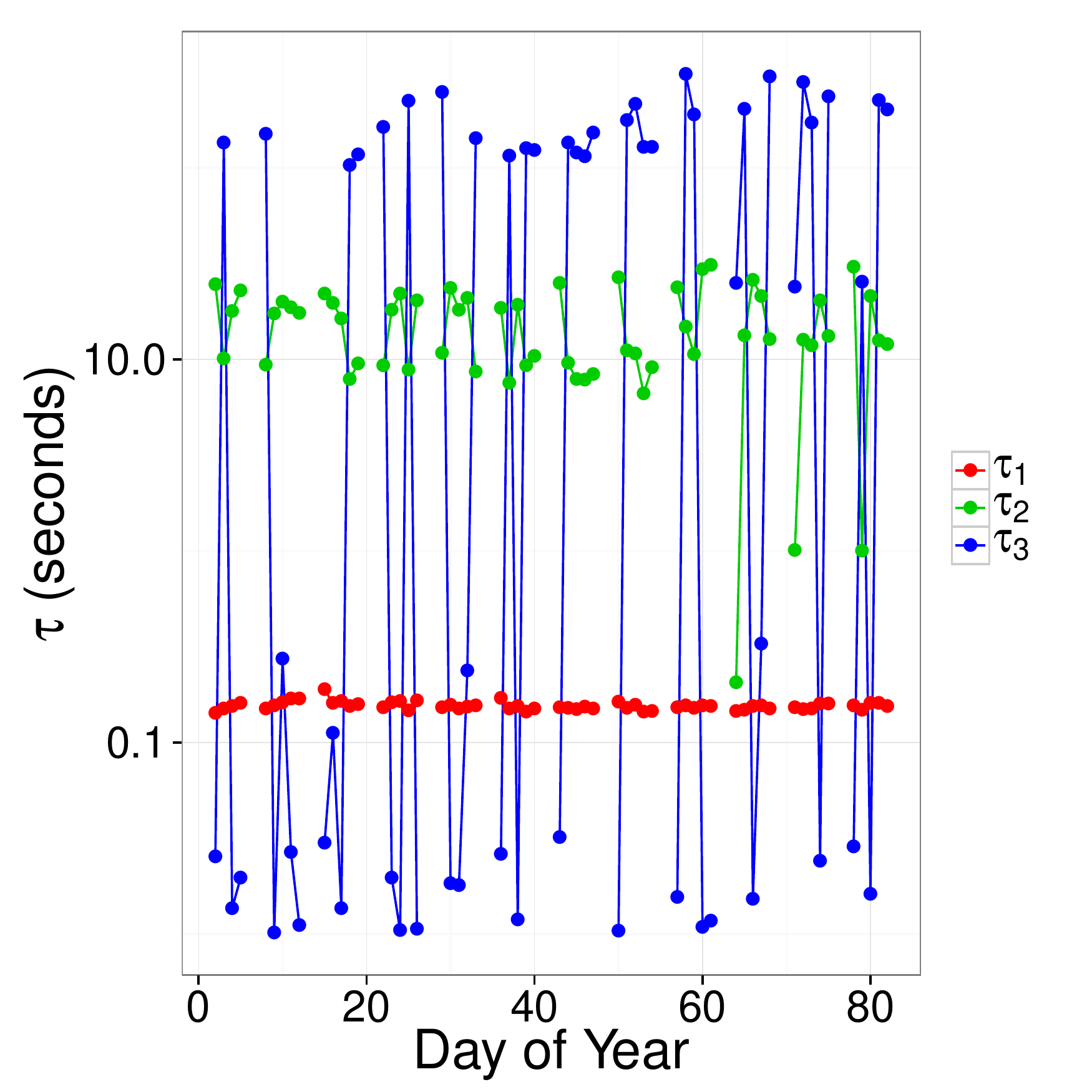}\protect\caption{\label{fig:Branching-ratio.}Daily branching ratio (left) and associated
time-scales (right). The shortest characteristic timescale is very
stable; the model captures $1$ or $2$ longer time scales depending
on the day.}
\end{figure}

While the total branching ratio oscillates around 0.8 (Fig.~\ref{fig:Branching-ratio}),
the parameters associated to each exponential make it clear that three
timescales are only found on some days. This, once again, may either
be because some days do not require three timescales, or because of
the sloppiness of sums of exponentials. As reported by Table~\ref{tab:Average-timescales-when},
the shortest timescale $\left\langle \tau_{1}\right\rangle $ does
not depend on the effective number of timescales, while the second
indeed does.

\begin{table}[H]

\begin{centering}
\begin{tabular}{|c|c|c|}
\hline 
 & 2 timescales & 3 timescales\tabularnewline
\hline 
\hline 
$\left\langle \tau_{1}\right\rangle $ & \unit[0.16]{s} & \unit[0.15]{s}\tabularnewline
\hline 
$\left\langle \tau_{2}\right\rangle $ & \unit[21.9]{s} &  \unit[9.3]{s}\tabularnewline
\hline 
$\left\langle \tau_{3}\right\rangle $ & NA & \unit[161]{s}\tabularnewline
\hline 
\end{tabular}\protect\caption{Average timescales when two or three timescales are found by fitting
$\phi_{3}$ to whole days.\label{tab:Average-timescales-when}}

\par\end{centering}

\end{table}

\subsection{Multi-day fits}

Extending fits to two days requires to account for weekly seasonality.
First and most importantly, EBS order book does not operate at week-ends,
which implies that Mondays and Fridays most likely have a dynamics
distinctly different from the other days. Thus we fit all pairs Tuesdays-Wednesdays,
and Wednesdays-Thursdays, which amounts to 26 fits (2 points per week,
13 weeks). Before proceeding, it is important to keep in mind that
Figure \ref{fig:Baseline-intensity-knots.} forewarns that the daily
variations of activity at various times of the day are ample, particularly
at about 4pm, the time of the daily fixing. This may also prevent
a single kernel to hold for several days in a row, the composition
of the reaction times of the population of traders being potentially
subject to similar fluctuations between two days.

\begin{table}[H]
\centering{}%
\begin{tabular}{|c|c|c|c|c|c|c|c|c|}
\hline 
Kernel & $\phi_{2}$ & $\phi_{3}$ & $\phi_{4}$ & $\phi_{15}^{{\scriptscriptstyle \text{HBB}}}$ & $\phi_{15}^{{\scriptscriptstyle \text{PL}}}$ & $\phi_{30}^{{\scriptscriptstyle \text{HBB}}}$ & $\phi_{30}^{{\scriptscriptstyle \text{PL}}}$ & $\phi_{15}^{{\scriptscriptstyle \text{PL}_{\text{x}}}}$\tabularnewline
\hline 
\hline 
$n$ & $0.80$ & $0.87$ & $0.88$ & $0.82$ & $0.84$ & $0.98$ & $0.97$ & $0.91$\tabularnewline
\hline 
$pKS$ & $0.02$ & $0.04$ & $0.06$ & $1\e{-12}$  & $3\e{-15}$  & $1\e{-6}$ & $3\e{-10}$ & $0.04$\tabularnewline
\hline 
$pED$ & $0.04$ & $0.46$ & $0.54$ & $0.50$ & $0.35$ & $0.59$ & $0.44$ & $0.58$\tabularnewline
\hline 
$pLB$ & $0.010$  & $0.008$  & $0.011$ & $2\e{-6}$  & $4\e{-6}$  & $3\e{-8}$  & $4\e{-8}$  & $0.001$\tabularnewline
\hline 
$\log\mathcal{L}_{p}$ & $119666.0$ & $119819.2$ & $119814.5$ & $119094.1$ & $119221.6$ & $119086.8$ & $119207.4$ & $119656.4$\tabularnewline
\hline 
$AIC_{p}$ & $-239303.5$ & $-239605.7$ & $-239592.4$ & $-238161.7$ & $-238416.7$ & $-238147.1$ & $-238388.2$ & $-239282.2$\tabularnewline
\hline 
$\epsilon$ & NA & NA & NA & $0.08$ & $0.11$ & $0.13$ & $0.15$ & $0.10$\tabularnewline
\hline 
$w$ & $0$ & $0.34$ & $0.40$ & $0$ & $0$ & $0$ & $0$ & $0.26$\tabularnewline
\hline 
$N_{max}$ & $0$ & $9$ & $10$ & $0$ & $0$ & $0$ & $0$ & $7$\tabularnewline
\hline 
\end{tabular}\protect\caption{Kernel comparison of two-days fits. 26 points.\label{tab:2days}}
\end{table}

Table \ref{tab:2days} compares the performance of all kernels. Kernel
$\phi_{2}$ performs poorly, while $\phi_{3}$, $\phi_{4}$ and $\phi_{15}^{{\scriptscriptstyle \text{PL}_{\text{x}}}}$
are the best ones according to $AIC_{p}$ criterion. No kernel can
pass the three tests at the same time ($\phi_{3}$ does for a single
pair of days). The timescales of $\phi_{3}$ are stable and similar
to those of single-day fits ( $\left\langle \tau_{1}\right\rangle \simeq$
\unit[0.15]{s}, $\left\langle \tau_{2}\right\rangle \simeq$ \unit[10.6]{s},
$\left\langle \tau_{3}\right\rangle \simeq$ \unit[178]{s}), while
$\phi_{4}$ sometimes manages to find a fourth timescale. For the
record, we tried to use 5 exponentials, but never found a fifth timescale.
It is noteworthy that $\phi_{4}$ has an acceptable average pKS. The
free exponential of $\phi_{15}^{{\scriptscriptstyle \text{PL}_{\text{x}}}}$
has a timescale of \unit[0.13]{s}.

\begin{figure}[H]
\centering{}\includegraphics[scale=0.35]{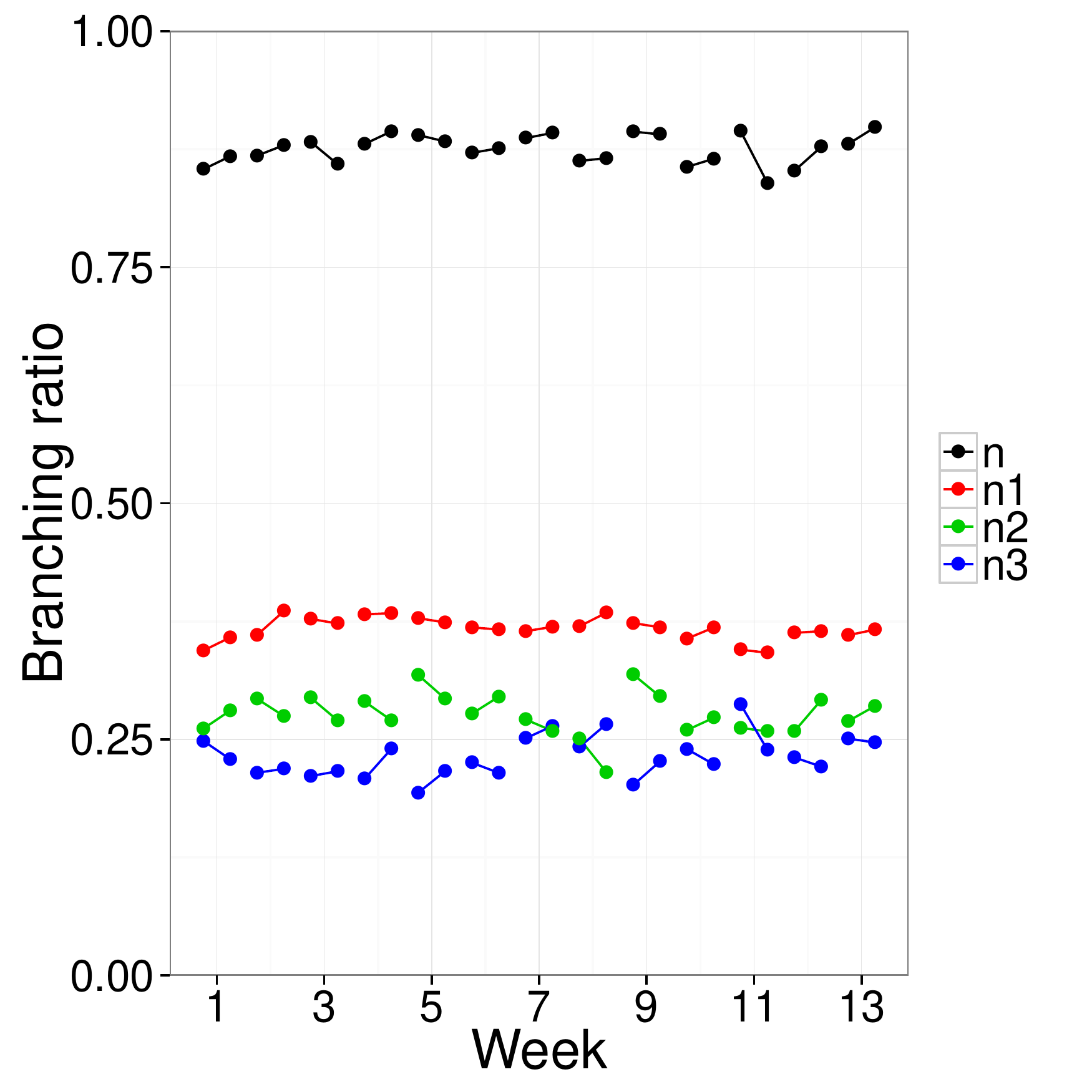}\includegraphics[scale=0.35]{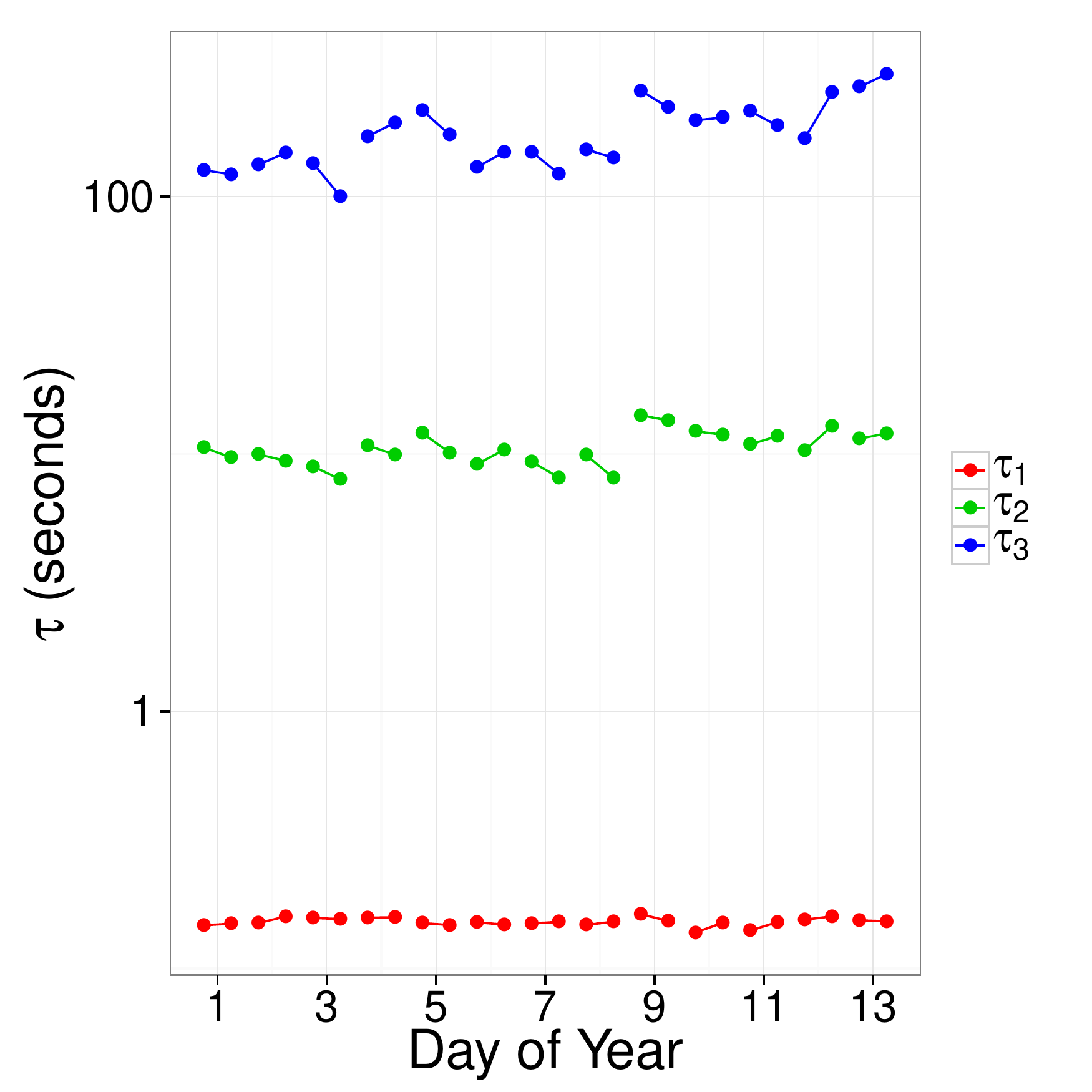}\protect\caption{Endogeneity factors (left plot) and associated timescales (right plot)
for fits of $\phi_{3}$ to two consecutive days\label{fig:2days_n}}
\end{figure}

\begin{table}[H]
\centering{}%
\begin{tabular}{|c|c|c|}
\hline 
 & 3 timescales & 4 timescales\tabularnewline
\hline 
\hline 
$\left\langle \tau_{1}\right\rangle $ & \unit[0.15]{s} & \unit[0.15]{s}\tabularnewline
\hline 
$\left\langle \tau_{2}\right\rangle $ & \unit[13.5]{s} &  \unit[7.1]{s}\tabularnewline
\hline 
$\left\langle \tau_{3}\right\rangle $ & \unit[226]{s} & \unit[33]{s}\tabularnewline
\hline 
$\left\langle \tau_{4}\right\rangle $ & NA & \unit[295]{s}\tabularnewline
\hline 
\end{tabular}\protect\caption{Average timescales when three or four timescales are found by fitting
$\phi_{4}$ to two consecutive days.\label{tab:Average-timescales-4}}
\end{table}

\section{Discussion and conclusions}

Our results are mostly positive: Hawkes processes can indeed be fitted
in a statistically significant way according to three tests to a whole
day of data. This means that they describe very precisely a large
number of events (around on average 15000). This is all the more remarkable
because the fitted timescales are quite small. This shows that the
endogenous part, which account for about 80\% of the events, is limited
to short time self-reactions in FX markets. This also means that at
these time horizons, the instantaneous distribution of reaction time
scales of the traders influences much the fitted kernels, as shown
by the lunch lull in endogeneity. This is one reason why fitting more
than one day with the same kernel is very hard since nothing guarantees
that the composition of the trader population will be the same for
several days in a row. 

Fitting longer and longer time periods requires more and more exponentials.
Fitting sums of exponentials with free parameters yields successive
timescales whose ratios are not constant, which contrasts with the
assumption of kernels that approximate power-laws. This is why the
kernel $\phi_{15}^{{\scriptscriptstyle \text{PL}_{\text{x}}}}$, which
adds one free exponential to the latter, has an overall better performance
than pure approximations of power-laws. Longer time periods also leads
to larger endogeneity factors, which makes sense since measuring long
memory by definition requires long time series. As it clearly appears
in all the tables, the use of power law-like kernels mechanically
increases the apparent endogeneity factor, some of them being dangerously
close to 1 (e.g $\phi_{30}^{{\scriptscriptstyle \text{HBB}}}$ and
$\phi_{30}^{{\scriptscriptstyle \text{PL}}}$). That said, and quite
importantly, the best kernels are never those with the largest endogeneity
factors. 

One may wonder if significance could be much improved by using data
with a much better time resolution. It would certainly help, but only
to a limited extent. As shown in Appendix \ref{sec:appendix_sims},
only the KS test is affected by introduction of limited time resolution.
Since the fits also fail to pass the the LB test for two consecutive
days that is not affected by a limited time resolution, it is safe
to assume that this failure has deeper reasons. The main problem resides
in the difficulties caused by the non-stationarities of both exogeneity
and endogeneity. The example of the lunch lull is striking: assuming
a constant kernel shape for all times of the day, while a good approximation,
cannot lead to statistical significance of fits over many days. In
this precise case, one could add a daily seasonality on some weights.

Our results may well be specific to FX markets. In particular, the
endogeneity is never close to 1, in contrast with studies on futures
on equity indices. However given the nightly closure of equities markets
(for example) and their short opening times, and given the difficulties
encountered for FX data, it seems difficult to envisage a statistically
satisfying comparison.

\appendix

\section{Simulations\label{sec:appendix_sims}}

We simulate a Hawkes process with a $\phi_{2}$ kernel with parameters
similar to those of hourly fits on real data: we set $\mu=0.05,$
$n_{1}=0.37$, $n_{2}=0.42$, $\tau_{2}=$\unit[21]{s} and vary $\tau_{1}$
from \unit[0.05]{s} to \unit[1.5]{s}. For each value of $\tau_{1}$
we perform $50$ simulations of $22$ hours. Then, on each simulated
time-series, we artificially reduce the data resolution to \unit[0.1]{s},
introducing time slicing as in our data set, and then randomize the
timestamps within each time slice in order to mimic the procedure
applied on empirical data (see Sec. \ref{sub:Treatment}). We fit
each resulting time-series and average the results over the $50$
runs with two- and three-exponential kernels. 

\begin{figure}[H]
\centering{}\includegraphics[scale=0.3]{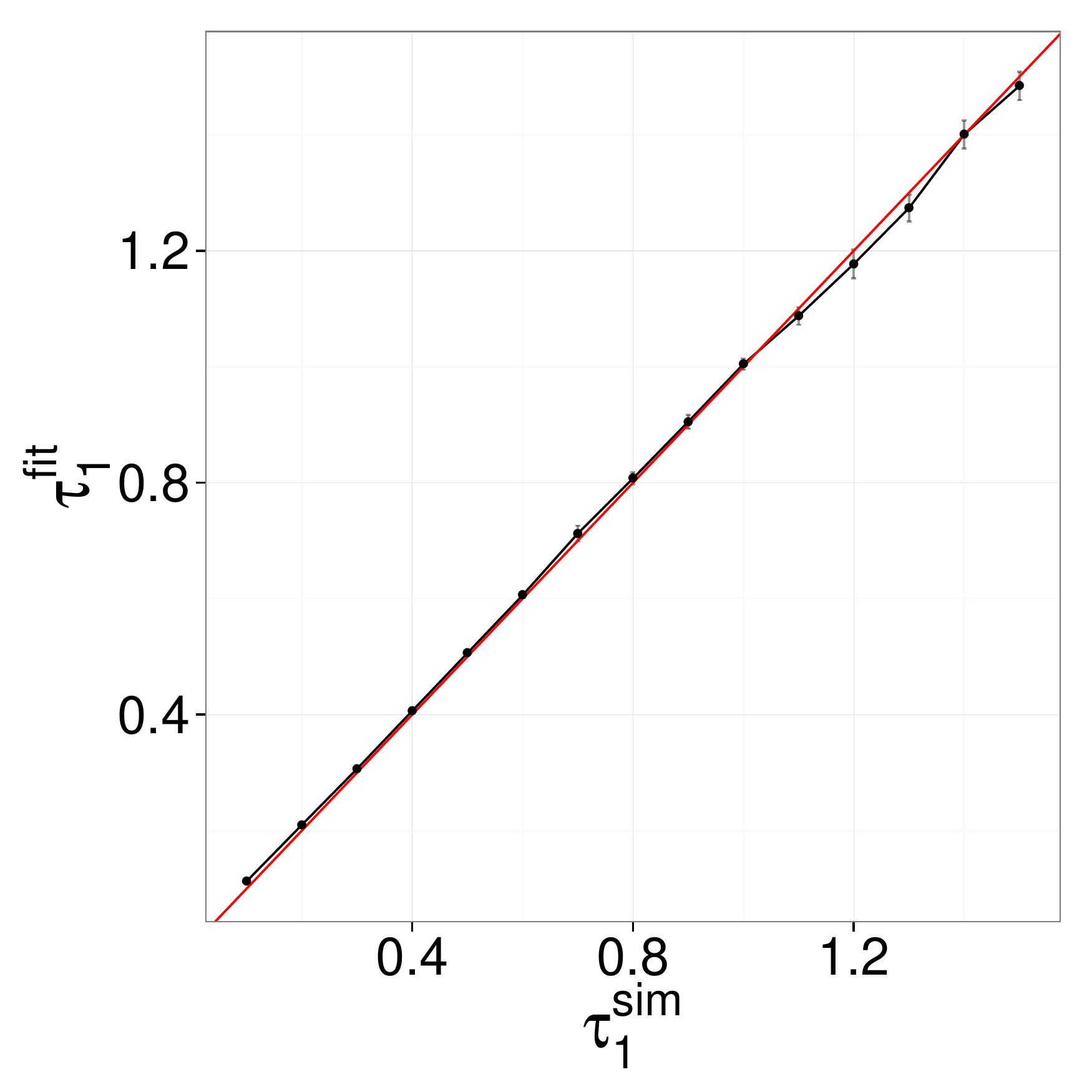}\includegraphics[scale=0.3]{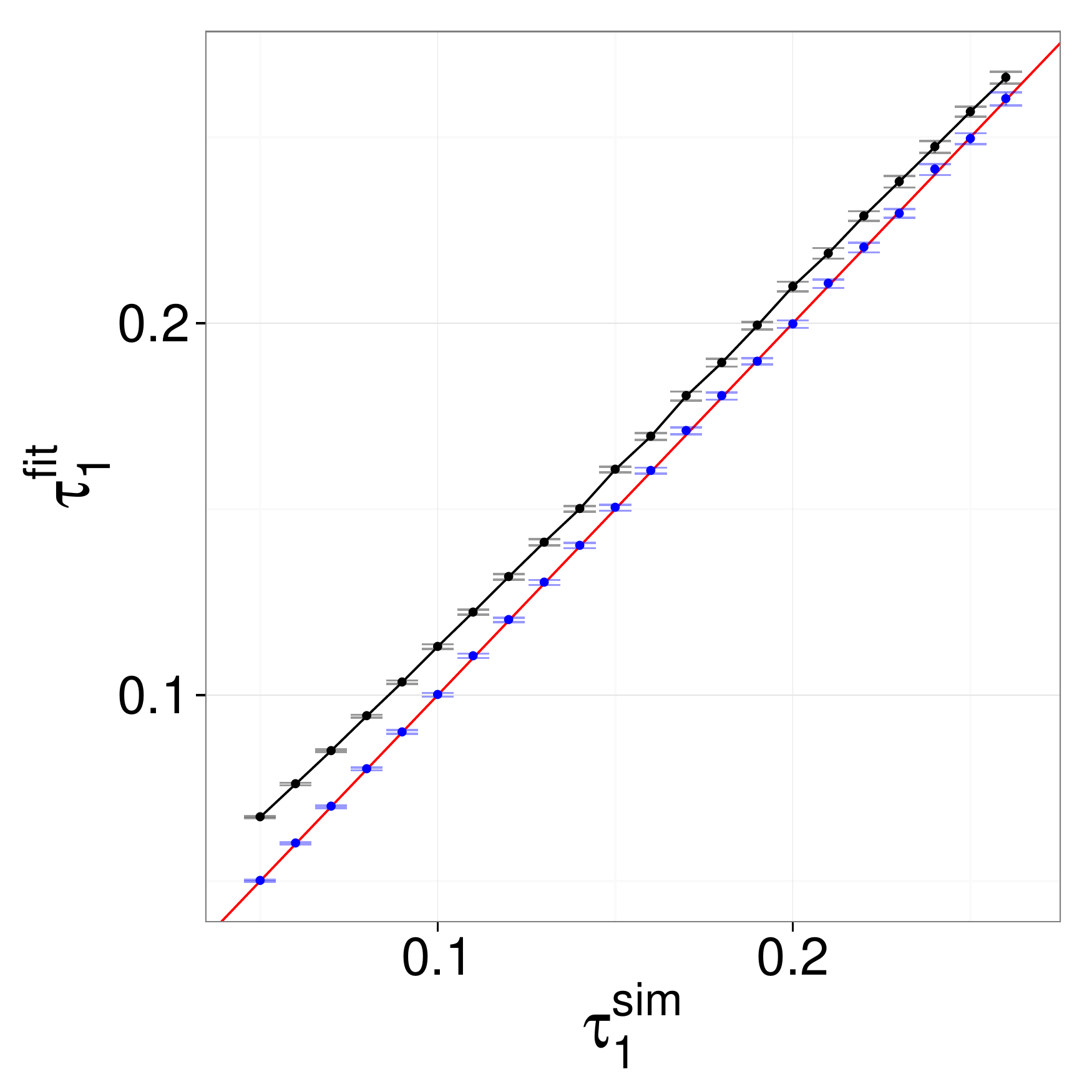}\protect\caption{Fitted short timescale (black points) versus simulated short timescale.
In red, the $y=x$ line. Right plot is a zoom on the critical region
(close to \unit[0.1]{s}). Blue points are the fitted values without
the slicing procedure. Small distortion in the short timescale determination.
Error bars set at two standard deviations.\label{fig:Fitted-short-timescale}. }
\end{figure}

Figure \ref{fig:Fitted-short-timescale} reports the fitted smallest
time scale as a function of the original time scale and shows that
the shuffling of time stamps within an interval leads artifically
increases the apparent smallest time scale, particularly (and quite
expectedly) for small $\tau_{1}^{\mbox{sim}}$. Nevertheless, this
increase is small, of the order of 15\%. In addition, shuffling does
not introduce a spurious third time scale, as fits with kernels with
three exponentials did not yield any third time scale.

\begin{figure}[H]
\begin{centering}
\includegraphics[scale=0.3]{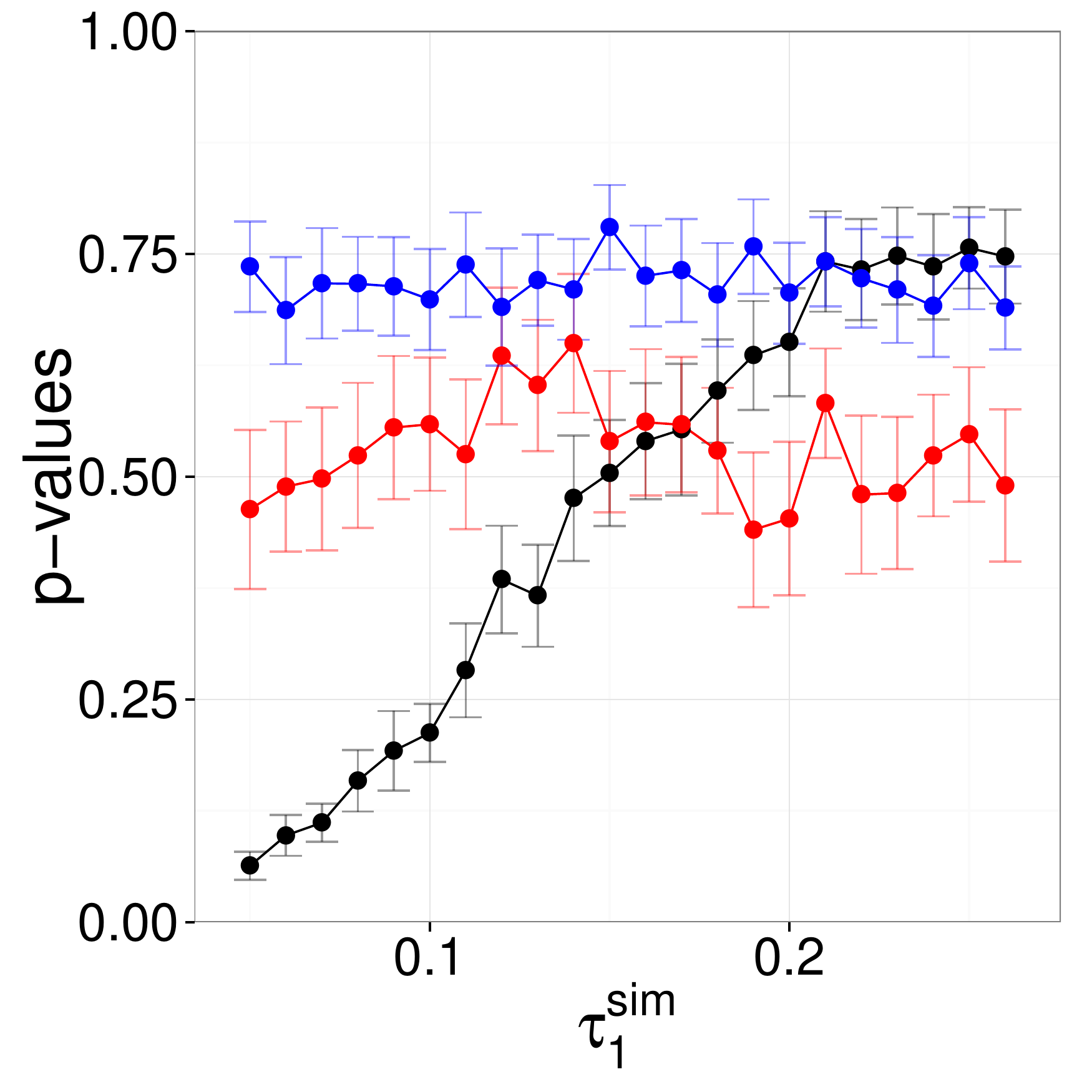}\includegraphics[scale=0.3]{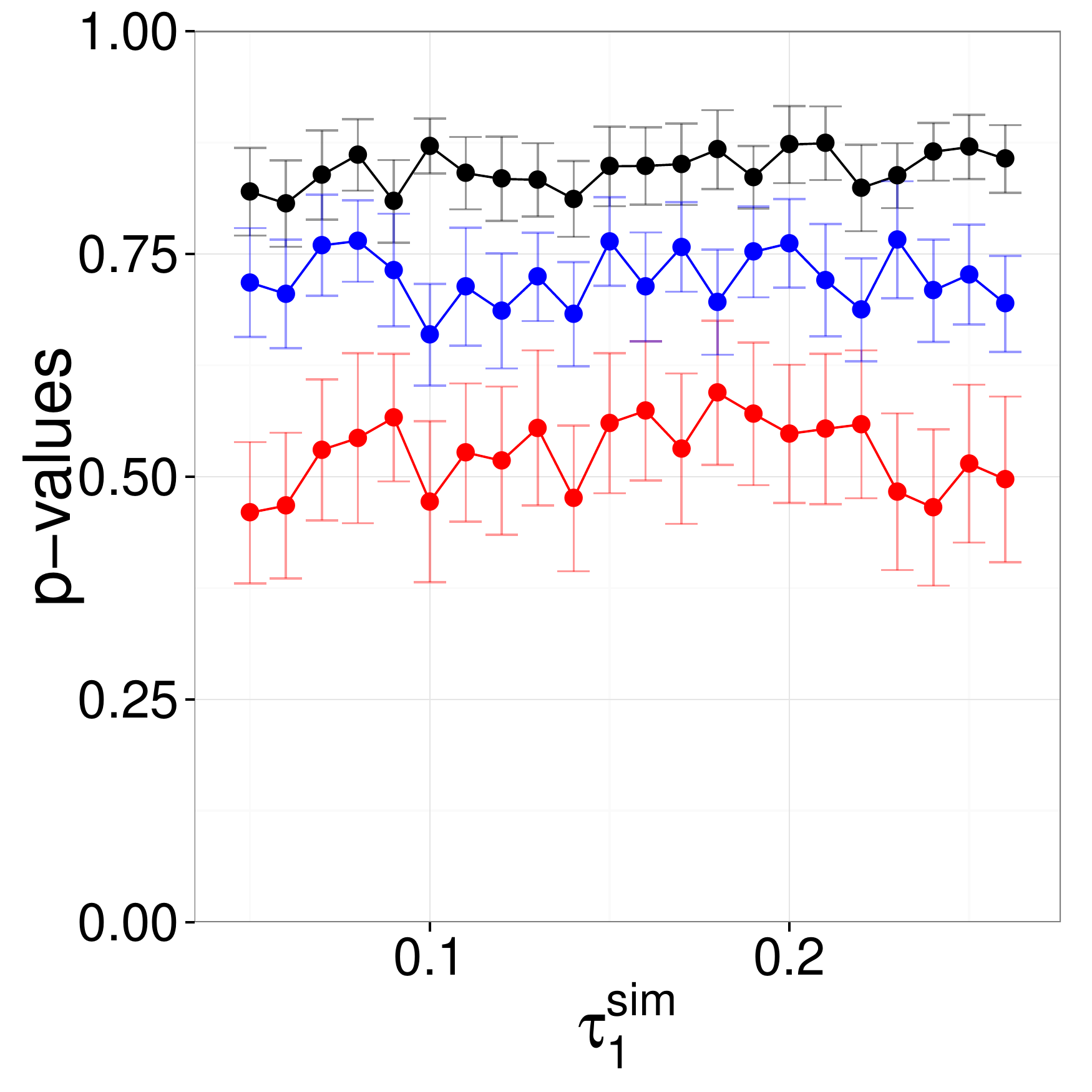}
\par\end{centering}

\protect\caption{Fits p-values for Kolmogorov-Smirnov test (black), Ljung-Box test
(red) and Excess-Dispersion test (blue). Left plot: with time stamp
shuffling within a time slice. Right plot: without shuffling. Only
the Kolmogorov-Smirnov p-value is affected by the data bundling. Error
bars set at two standard deviations. \label{fig:Fits-p-values-for}}

\end{figure}

Finally, Fig.~\ref{fig:Fits-p-values-for} shows that only the Kolmogorov-Smirnov
p-values are affected by the time slicing and time stamp shuffling
within a time slice. Nevertheless, at $\tau_{1}^{fit}=0.15\mbox{s}$
, $pKS$ is still larger than $0.05.$ This is consistent with fits
on real data: significance is possible, but limited time resolution
does not help. 

\bibliographystyle{plainnat}
\bibliography{Hawkes}

\begin{thebibliography}{45}
\providecommand{\natexlab}[1]{#1}
\providecommand{\url}[1]{\texttt{#1}}
\expandafter\ifx\csname urlstyle\endcsname\relax
  \providecommand{\doi}[1]{doi: #1}\else
  \providecommand{\doi}{doi: \begingroup \urlstyle{rm}\Url}\fi

\bibitem[A\"{\i}t-Sahalia et~al.(2010)A\"{\i}t-Sahalia, Cacho-Diaz, and
  Laeven]{Ait-Sahalia2010}
Yacine A\"{\i}t-Sahalia, Julio Cacho-Diaz, and Roger J~A Laeven.
\newblock {Modeling Financial Contagion Using Mutually Exciting Jump
  Processes}.
\newblock \emph{National Bureau of Economic Research Working Paper Series}, No.
  15850, 2010.

\bibitem[Bacry et~al.(2012{\natexlab{a}})Bacry, Dayri, and Muzy]{Bacry2012}
E.~Bacry, K.~Dayri, and J.~F. Muzy.
\newblock {Non-parametric kernel estimation for symmetric Hawkes processes.
  Application to high frequency financial data}.
\newblock \emph{The European Physical Journal B}, 85\penalty0 (5):\penalty0
  157, May 2012{\natexlab{a}}.

\bibitem[Bacry et~al.(2012{\natexlab{b}})Bacry, Delattre, Hoffmann, and
  Muzy]{Bacry2012a}
E~Bacry, S~Delattre, M~Hoffmann, and J~F Muzy.
\newblock {Modelling microstructure noise with mutually exciting point
  processes}.
\newblock \emph{Quantitative Finance}, 13\penalty0 (1):\penalty0 65--77,
  January 2012{\natexlab{b}}.

\bibitem[Bacry and Muzy(2014{\natexlab{a}})]{Bacry2014a}
Emmanuel Bacry and Jean-Fran\c{c}ois Muzy.
\newblock {Hawkes model for price and trades high-frequency dynamics}.
\newblock \emph{Quantitative Finance}, pages 1--20, April 2014{\natexlab{a}}.

\bibitem[Bacry and Muzy(2014{\natexlab{b}})]{Bacry2014}
Emmanuel Bacry and Jean-Francois Muzy.
\newblock {Second order statistics characterization of Hawkes processes and
  non-parametric estimation}.
\newblock January 2014{\natexlab{b}}.

\bibitem[Bauwens and Hautsch(2003)]{Bauwens2004}
Luc Bauwens and Nikolaus Hautsch.
\newblock {Dynamic Latent Factor Models for Intensity Processes}.
\newblock \emph{CORE Discussion Paper}, 103, February 2003.

\bibitem[Bormetti et~al.(2013)Bormetti, Calcagnile, Treccani, Corsi, Marmi, and
  Lillo]{Bormetti2013}
Giacomo Bormetti, Lucio~Maria Calcagnile, Michele Treccani, Fulvio Corsi,
  Stefano Marmi, and Fabrizio Lillo.
\newblock {Modelling systemic price cojumps with Hawkes factor models}.
\newblock January 2013.

\bibitem[Bowsher(2007)]{Bowsher2007}
Clive~G Bowsher.
\newblock {Modelling security market events in continuous time: Intensity
  based, multivariate point process models}.
\newblock \emph{Journal of Econometrics}, 141\penalty0 (2):\penalty0 876--912,
  December 2007.

\bibitem[Br\'{e}maud and Massouli\'{e}(2001)]{Bremaud2001}
Pierre Br\'{e}maud and Laurent Massouli\'{e}.
\newblock {Hawkes Branching Point Processes without Ancestors}.
\newblock \emph{Journal of Applied Probability}, 38\penalty0 (1):\penalty0
  122--135, March 2001.

\bibitem[Byrd et~al.(1995)Byrd, Lu, Nocedal, and Zhu]{Byrd1995}
R~Byrd, P~Lu, J~Nocedal, and C~Zhu.
\newblock {A Limited Memory Algorithm for Bound Constrained Optimization}.
\newblock \emph{SIAM Journal on Scientific Computing}, 16\penalty0
  (5):\penalty0 1190--1208, September 1995.

\bibitem[Chavez-Demoulin and McGill(2012)]{Chavez-Demoulin2012}
V~Chavez-Demoulin and J~A McGill.
\newblock {High-frequency financial data modeling using Hawkes processes}.
\newblock \emph{Journal of Banking \& Finance}, 36\penalty0 (12):\penalty0
  3415--3426, December 2012.

\bibitem[Chavez-Demoulin et~al.(2005)Chavez-Demoulin, Davison, and
  McNeil]{Chavez-Demoulin*2005}
V~Chavez-Demoulin, A~C Davison, and A~J McNeil.
\newblock {Estimating value-at-risk: a point process approach}.
\newblock \emph{Quantitative Finance}, 5\penalty0 (2):\penalty0 227--234, April
  2005.

\bibitem[Chornoboy et~al.(1988)Chornoboy, Schramm, and Karr]{Chornoboy1988}
E~S Chornoboy, L~P Schramm, and A~F Karr.
\newblock {Maximum likelihood identification of neural point process systems}.
\newblock \emph{Biological Cybernetics}, 59\penalty0 (4-5):\penalty0 265--275,
  1988.

\bibitem[Crane and Sornette(2008)]{Crane2008}
Riley Crane and Didier Sornette.
\newblock {Robust dynamic classes revealed by measuring the response function
  of a social system}.
\newblock \emph{Proceedings of the National Academy of Sciences}, 105\penalty0
  (41):\penalty0 15649--15653, October 2008.

\bibitem[Cutler et~al.(1989)Cutler, Poterba, and Summers]{cutler1989moves}
David~M Cutler, James~M Poterba, and Lawrence~H Summers.
\newblock What moves stock prices?
\newblock \emph{The Journal of Portfolio Management}, 15\penalty0 (3):\penalty0
  4--12, 1989.

\bibitem[{Da Fonseca} and Zaatour(2013)]{DaFonseca2013}
Jos\'{e} {Da Fonseca} and Riadh Zaatour.
\newblock {Hawkes process: Fast calibration, application to trade clustering,
  and diffusive limit}.
\newblock \emph{Journal of Futures Markets}, pages n/a--n/a, 2013.

\bibitem[Dacorogna et~al.(1993)Dacorogna, M\"{u}ller, Nagler, Olsen, and
  Pictet]{Dacorogna1993}
Michael~M Dacorogna, Ulrich~A M\"{u}ller, Robert~J Nagler, Richard~B Olsen, and
  Olivier~V Pictet.
\newblock {A geographical model for the daily and weekly seasonal volatility in
  the foreign exchange market}.
\newblock \emph{Journal of International Money and Finance}, 12\penalty0
  (4):\penalty0 413--438, August 1993.

\bibitem[Engle and Russell(1998)]{Engle1998}
Robert~F Engle and Jeffrey~R Russell.
\newblock {Autoregressive Conditional Duration: A New Model for Irregularly
  Spaced Transaction Data}.
\newblock \emph{Econometrica}, 66\penalty0 (5):\penalty0 1127--1162, September
  1998.

\bibitem[Errais et~al.(2010)Errais, Giesecke, and Goldberg]{Errais2010}
E~Errais, K~Giesecke, and L~Goldberg.
\newblock {Affine Point Processes and Portfolio Credit Risk}.
\newblock \emph{SIAM Journal on Financial Mathematics}, 1\penalty0
  (1):\penalty0 642--665, January 2010.

\bibitem[Filimonov and Sornette(2012)]{Filimonov2012}
Vladimir Filimonov and Didier Sornette.
\newblock {Quantifying reflexivity in financial markets: Toward a prediction of
  flash crashes}.
\newblock \emph{Physical Review E}, 85\penalty0 (5):\penalty0 056108, May 2012.

\bibitem[Filimonov and Sornette(2013)]{Filimonov2013}
Vladimir Filimonov and Didier Sornette.
\newblock {Apparent criticality and calibration issues in the Hawkes
  self-excited point process model: application to high-frequency financial
  data}.
\newblock August 2013.

\bibitem[Hardiman et~al.(2013)Hardiman, Bercot, and Bouchaud]{Hardiman2013a}
Stephen~J Hardiman, Nicolas Bercot, and Jean-Philippe Bouchaud.
\newblock {Critical reflexivity in financial markets: a Hawkes process
  analysis}.
\newblock \emph{Eur. Phys. J. B}, 86\penalty0 (10), October 2013.

\bibitem[Hawkes(1971{\natexlab{a}})]{Hawkes1971}
Alan~G Hawkes.
\newblock {Spectra of Some Self-Exciting and Mutually Exciting Point
  Processes}.
\newblock \emph{Biometrika}, 58\penalty0 (1):\penalty0 83--90, April
  1971{\natexlab{a}}.

\bibitem[Hawkes(1971{\natexlab{b}})]{Hawkes1971a}
Alan~G Hawkes.
\newblock {Point Spectra of Some Mutually Exciting Point Processes}.
\newblock \emph{Journal of the Royal Statistical Society. Series B
  (Methodological)}, 33\penalty0 (3):\penalty0 438--443, January
  1971{\natexlab{b}}.

\bibitem[Hewlett(2006)]{Hewlett2006}
Patrick Hewlett.
\newblock {Clustering of order arrivals, price impact and trade path
  optimisation}.
\newblock In \emph{Workshop on Financial Modeling with Jump Processes}, 2006.

\bibitem[Ito and Hashimoto(2006)]{Ito2006}
Takatoshi Ito and Yuko Hashimoto.
\newblock {Intraday seasonality in activities of the foreign exchange markets:
  Evidence from the electronic broking system}.
\newblock \emph{Journal of the Japanese and International Economies},
  20\penalty0 (4):\penalty0 637--664, December 2006.

\bibitem[Jaisson and Rosenbaum(2013)]{Jaisson2013}
Thibault Jaisson and Mathieu Rosenbaum.
\newblock {Limit theorems for nearly unstable Hawkes processes}.
\newblock page~38, October 2013.

\bibitem[Jedidi and Abergel(2013)]{Jedidi2013}
Aymen Jedidi and Frederic Abergel.
\newblock {On the Stability and Price Scaling Limit of a Hawkes Process-Based
  Order Book Model}.
\newblock \emph{SSRN Electronic Journal}, May 2013.

\bibitem[Joulin et~al.(2008)Joulin, Lefevre, Grunberg, and
  Bouchaud]{joulin2008stock}
Armand Joulin, Augustin Lefevre, Daniel Grunberg, and Jean-Philippe Bouchaud.
\newblock Stock price jumps: news and volume play a minor role.
\newblock \emph{arXiv preprint arXiv:0803.1769}, 2008.

\bibitem[Large(2007)]{Large2007}
Jeremy Large.
\newblock {Measuring the resiliency of an electronic limit order book}.
\newblock \emph{Journal of Financial Markets}, 10\penalty0 (1):\penalty0 1--25,
  February 2007.

\bibitem[Lewis and Mohler(2011)]{Lewis2011}
E~Lewis and G~Mohler.
\newblock {A Nonparametric EM algorithm for Multiscale Hawkes Processes}.
\newblock \emph{Submitted}, 2011.

\bibitem[Ljung and Box(1978)]{LJUNG1978}
G~M Ljung and G~E~P Box.
\newblock {On a measure of lack of fit in time series models}.
\newblock \emph{Biometrika}, 65\penalty0 (2):\penalty0 297--303, August 1978.

\bibitem[Marsan and Lenglin\'{e}(2008)]{Marsan2008}
David Marsan and Olivier Lenglin\'{e}.
\newblock {Extending Earthquakes' Reach Through Cascading}.
\newblock \emph{Science}, 319\penalty0 (5866):\penalty0 1076--1079, February
  2008.

\bibitem[Mohler et~al.(2011)Mohler, Short, Brantingham, Schoenberg, and
  Tita]{Mohler2011}
G~O Mohler, M~B Short, P~J Brantingham, F~P Schoenberg, and G~E Tita.
\newblock {Self-Exciting Point Process Modeling of Crime}.
\newblock \emph{Journal of the American Statistical Association}, 106\penalty0
  (493):\penalty0 100--108, March 2011.

\bibitem[Ogata(1999)]{Ogata1999}
Y~Ogata.
\newblock {Seismicity Analysis through Point-process Modeling: A Review}.
\newblock \emph{pure and applied geophysics}, 155\penalty0 (2-4):\penalty0
  471--507, 1999.

\bibitem[Ogata(1988)]{Ogata1988}
Yosihiko Ogata.
\newblock {Statistical Models for Earthquake Occurrences and Residual Analysis
  for Point Processes}.
\newblock \emph{Journal of the American Statistical Association}, 83\penalty0
  (401):\penalty0 9--27, March 1988.

\bibitem[Ozaki(1979)]{Ozaki1979}
T~Ozaki.
\newblock {Maximum likelihood estimation of Hawkes' self-exciting point
  processes}.
\newblock \emph{Annals of the Institute of Statistical Mathematics},
  31\penalty0 (1):\penalty0 145--155, 1979.

\bibitem[Pernice et~al.(2012)Pernice, Staude, Cardanobile, and
  Rotter]{Pernice2012}
Volker Pernice, Benjamin Staude, Stefano Cardanobile, and Stefan Rotter.
\newblock {Recurrent interactions in spiking networks with arbitrary topology}.
\newblock \emph{Physical Review E}, 85\penalty0 (3):\penalty0 31916, March
  2012.

\bibitem[Rambaldi et~al.(2014)Rambaldi, Pennesi, and Lillo]{Rambaldi2014}
Marcello Rambaldi, Paris Pennesi, and Fabrizio Lillo.
\newblock {Modeling FX market activity around macroeconomic news: a Hawkes
  process approach}.
\newblock page~11, May 2014.

\bibitem[Reynaud-Bouret and Schbath(2010)]{Reynaud-Bouret2010}
Patricia Reynaud-Bouret and Sophie Schbath.
\newblock {Adaptive estimation for Hawkes processes; application to genome
  analysis}.
\newblock pages 2781--2822, 2010.

\bibitem[Soros(1987)]{Soros1987}
Georges Soros.
\newblock \emph{{The Alchemy of Finance: Reding the Mind of the Market}}.
\newblock 1987.

\bibitem[Toke and Pomponio(2011)]{Toke2011}
Ioane~Muni Toke and Fabrizio Pomponio.
\newblock {Modelling Trades-Through in a Limited Order Book Using Hawkes
  Processes Trades-through}.
\newblock 2011.

\bibitem[Wagenmakers and Farrell(2004)]{Wagenmakers2004}
Eric-Jan Wagenmakers and Simon Farrell.
\newblock {AIC model selection using Akaike weights}.
\newblock \emph{Psychonomic Bulletin \& Review}, 11\penalty0 (1):\penalty0
  192--196, 2004.

\bibitem[Waterfall et~al.(2006)Waterfall, Casey, Gutenkunst, Brown, Myers,
  Brouwer, Elser, and Sethna]{Waterfall2006}
Joshua~J Waterfall, Fergal~P Casey, Ryan~N Gutenkunst, Kevin~S Brown,
  Christopher~R Myers, Piet~W Brouwer, Veit Elser, and James~P Sethna.
\newblock {Sloppy-Model Universality Class and the Vandermonde Matrix}.
\newblock \emph{Physical Review Letters}, 97\penalty0 (15):\penalty0 150601,
  October 2006.

\bibitem[Yang and Zha(2013)]{Yang2013}
S.~Yang and H.~Zha.
\newblock {Mixture of Mutually Exciting Processes for Viral Diffusion}.
\newblock \emph{Journal of Machine Learning Research}, 28\penalty0
  (2):\penalty0 1--9, 2013.

\end{thebibliography}

\end{document}